\documentclass[%
 reprint,
 amsmath,amssymb,
 aps,
]{revtex4-2}

\usepackage{amsfonts,amsmath,amssymb,mathrsfs,amsthm}
\usepackage{epsfig,epstopdf,graphicx,graphics}
\usepackage{color}
\usepackage{float}
\usepackage{ifpdf}
\usepackage{braket}
\usepackage{algorithm}
\usepackage{algorithmic}
\usepackage{dcolumn}
\usepackage{bm}
\usepackage{tikz}
\usepackage{tikz-qtree}
\usepackage{hyperref}

\newtheorem{theorem}{Theorem}
\newtheorem{lemma}{Lemma}
\newcommand{\bthm}{\begin{theorem}}
\newcommand{\ethm}{\end{theorem}}
\newcommand{\blem}{\begin{lemma}}
\newcommand{\elem}{\end{lemma}}
\newcommand{\bex}{\begin{example}}
\newcommand{\eex}{\end{example}}
\newcommand{\bprop}{\begin{proposition}}
\newcommand{\eprop}{\end{proposition}}
\newcommand{\bplm}{\begin{problem}}
\newcommand{\eplm}{\end{problem}}
\newcommand{\bmrk}{\begin{remark}}
\newcommand{\emrk}{\end{remark}}
\newcommand{\bdfn}{\begin{definition}}
\newcommand{\edfn}{\end{definition}}
\newcommand{\bcor}{\begin{corollary}}
\newcommand{\ecor}{\end{corollary}}

\newcommand{\beq}{\begin{equation}}
\newcommand{\eeq}{\end{equation}}
\newcommand{\beqm}{\begin{equation*}}
\newcommand{\eeqm}{\end{equation*}}
\newcommand{\beqn}{\begin{eqnarray}}
\newcommand{\eeqn}{\end{eqnarray}}
\newcommand{\beqnm}{\begin{eqnarray*}}
\newcommand{\eeqnm}{\end{eqnarray*}}
\newcommand{\bea}{\begin{align}}
\newcommand{\eea}{\end{align}}
\newcommand{\beam}{\begin{align*}}
\newcommand{\eeam}{\end{align*}}
\newcommand{\bei}{\begin{itemize}}
\newcommand{\eei}{\end{itemize}}
\newcommand{\bed}{\begin{description}}
\newcommand{\eed}{\end{description}}
\newcommand{\bee}{\begin{enumerate}}
\newcommand{\eee}{\end{enumerate}}
\newcommand{\bey}{\begin{array}}
\newcommand{\eey}{\end{array}}
\newcommand{\beb}{\begin{thebibliography}{99}}
\newcommand{\eeb}{\end{thebibliography}}
\newcommand{\mbb}{\mathbb}

\newcommand{\A}{{\cal A}}
\newcommand{\BA}{{\bf A}}
\newcommand{\B}{{\cal B}}
\newcommand{\C}{{\bf C}}
\newcommand{\MI}{{\bf I}}
\newcommand{\T}{{\cal T}}
\newcommand{\BS}{{\cal S}}
\newcommand{\Y}{{\cal Y}}

\newcommand{\M}{{\bf M}}
\newcommand{\MP}{{\bf P}}
\newcommand{\MQ}{{\bf Q}}
\newcommand{\U}{{\bf U}}
\newcommand{\W}{{\bf W}}

\newcommand{\x}{{\bf x}}

\newcommand{\ii}{{\rm i}}
\def\l{\langle}
\def\r{\rangle}
\def\ff{\frac}

\begin{document}

\preprint{APS/123-QED}

\title{Quantum Higher Order Singular Value Decomposition}

\author{Lejia Gu}
 \altaffiliation{le-jia.gu@connect.polyu.hk}
\author{Xiaoqiang Wang}%
 \email{xiaoqiang.wang@connect.polyu.hk}
\author{H. W. Joseph Lee}%
 \email{joseph.lee@polyu.edu.hk}
\author{Guofeng Zhang}
 \homepage{guofeng.zhang@polyu.edu.hk}
\affiliation{%
 Department of Applied Mathematics, The Hong Kong Polytechnic University, Hong Kong\\
}%





\date{\today}

\begin{abstract}
Higher order singular value decomposition (HOSVD) is an important tool for analyzing big data in multilinear algebra and machine learning. In this paper, we present two quantum algorithms for HOSVD. Our methods allow one to decompose a tensor into a core tensor containing tensor singular values and some unitary matrices by quantum computers. Compared to the classical HOSVD algorithm, our quantum algorithms provide an exponential speedup. Furthermore, we introduce a hybrid quantum-classical algorithm of HOSVD model applied in recommendation systems.
\end{abstract}


\maketitle


\section{Introduction}

Matrix computations are vital to many optimization and machine learning problems. Nowadays, due to the rise of neural networks in machine learning methods, the elements of a network are usually described by tensors which can have more than two indices. Tensors (or hypermatrices), as a higher order generalization of matrices, have found widespread applications in scientific and engineering fields \cite{KB09, TTD, QCC, Qi}. Tensor decomposition describes a tensor as a sequence of elementary operations acting on other, often simpler tensors. Usually, key information can be extracted from the decomposed tensor, and less space is required to store the original tensor to some accuracy. Tensor network, as a countable collection of tensors connected by contractions, has been widely employed in training machine learning models. A quantum state has a tensor representation. Hence, a quantum network, namely a multipartite system, can be represented by a tensor network. Indeed, quantum circuits are a special class of tensor networks, where the arrangement of the tensors and their types are restricted \cite{QTN,MLQD,TQML}. Moreover, tensor analysis has been applied to the quantum entanglement problem and classicality problem of spin states \cite{HQZ16,QZN18,ZNZ19,QZB17}.

Quantum computers are devices that perform calculations by utilizing quantum mechanical features including superposition and entanglement. Although large-scale quantum computers are not built yet, theoretical research on quantum algorithms has been conducted for a few decades. In 1994, Shor's algorithm \cite{Shor}, a hybrid quantum-classical algorithm, is proved to be able to solve integer-factorization problem with polynomial time, while it is widely regarded as an NP hard problem in classical computing. In 1996, Grover's search algorithm \cite{Grover} finds an entry from an unstructured database quadratically faster than classical algorithms. In 2009, Harrow, Hassidim and Lloyd put forward a quantum algorithm for solving linear systems of equations, which is famous as the HHL Algorithm \cite{HHL}. Physical demonstrations of the HHL algorithm can be found in \cite{CWS+13,PCY+14,ZPR19}. Base on this algorithm, many quantum versions of classical machine learning methods are designed, such as quantum least-squares linear regression \cite{QDT} and support vector machines \cite{QSVM}. Also, the quantum singular value decomposition of nonsparse low-rank matrices is introduced in \cite{QSVD}. A different method to implement singular value decomposition, named as quantum singular value estimation (QSVE), is proposed in \cite{QRS}. The runtimes of the mentioned quantum algorithms are both polylogarithmic in the dimensions of the matrix, so they provide exponential speedups over their classical counterparts.

There are several types of tensor decompositions, such as CANDECOMP/PARAFAC (CP) decomposition \cite{CP1,CP2}, tensor-train (TT) decomposition \cite{TTD}, Tucker decomposition \cite{TD}, and etc. However, currently there are no quantum tensor decomposition algorithms. In this paper, we propose a quantum higher order singular value decomposition (Q-HOSVD). HOSVD is a specific orthogonal Tucker decomposition, and can be considered as an extension of SVD from matrices to tensors.

Classical HOSVD has been well studied, see, e.g., De Lathauwer, De Moor, and Vandewalle in 2000 \cite{Lie2000}, and it has been successfully applied to signal processing \cite{DNA} and pattern recognition \cite{image} problems. Furthermore, HOSVD has shown its strong power in quantum chemistry, especially in the second order M{\o}ller Plesset perturbation theory calculations \cite{QCh}. In addition, HOSVD is used in \cite{Z17} to derive the output $m$ photon state of a quantum linear passive system which is driven by an $m$ photon input state; more specifically, the wave function of the output is expressed in terms of the HOSVD of the input wave function.

Since HOSVD deals with high dimensional data, it has been put into practice in some machine learning methods. For example, it has been successfully applied in recommendation systems \cite{Mulre,Tagre}. In \cite{NN}, HOSVD representation for neural networks is proposed. By applying HOSVD the parameter-varying system can be expressed in a tensor product form by locally tuned neural network models. Additionally in \cite{CNN}, HOSVD is applied for compressing convolutional neural networks (CNN).

Our Q-HOSVD algorithms are based upon the quantum matrix singular value decomposition algorithm \cite{QSVD}, quantum singular value estimation \cite{QRS}, and several other quantum computing techniques. The input  can be a tensor of any order and dimension. By our Q-HOSVD algorithms, it is possible to perform singular value decomposition on tensors exponentially faster than classical algorithms. It can be directly applied to quantum machine learning algorithms, and may help solve computationally challenging problems arising in quantum mechanics and chemistry.

Compared to our preliminary research \cite{GWZ19}, the current paper provides two different algorithms to implement quantum HOSVD. Algorithm 1 is similar to that in  \cite{GWZ19}; however, here we present more details on the derivations, and in particular build the SWAP operator for quantum principal component analysis (qPCA) in a direct way.  An alternative quantum HOSVD algorithm, Algorithm 2, is proposed in the current paper which is based on quantum singular value estimation \cite{QRS}. Moreover, a thorough explanation for the computational complexity of these two algorithms is presented. Finally, we propose a {\it hybrid} quantum-classical algorithm of HOSVD model applied in recommendation systems.

The remainder of this paper is organized as follows. Some preliminaries are given in Section \ref{pre}. Two quantum higher order singular value decomposition algorithms are presented in Sections \ref{a1} and \ref{a2} respectively. The computational complexity is discussed in Section \ref{comp}. In Section \ref{rs}, we give an application of HOSVD model on quantum recommendation systems. In the last section, we summarize the results and compare the quantum HOSVD algorithm with the classical counterpart.

\section{Preliminaries}\label{pre}
First, we would like to add a comment on the notation that is used. Different symbols are used to facilitate the distinction among scalars, vectors, matrices, and tensors. Scalars are denoted by both lower-case letters $(a, b, \ldots ; \alpha, \beta, \ldots)$ and capital letters $(A,B,\ldots)$. Bold-face lower-case letters $(\bf{a}, \bf{b},\ldots)$ represent vectors. Since the algorithms we present are quantum algorithms, vectors are represented as quantum states after this section, ket $\ket{\cdot}$ denotes a column vector, and bra $\bra{\cdot}$ denotes a row vector. Bold-face capitals $(\bf{A}, \bf{B}, \ldots )$ correspond to matrices or operators, and tensors are written as calligraphic letters $(\A, \B, \ldots )$.

For a matrix ${\bf M}\in\mathbb{C}^{m\times n}$, its singular value decomposition (SVD) reads
\beq
{\bf M}={\bf U\Sigma V}^\dagger,
\eeq
where $\bf U$ is an $m \times m$ unitary matrix, $\bf \Sigma$ is a diagonal $m \times n$ matrix with non-negative real values on the diagonal, $\bf V$ is an $n \times n$ unitary matrix and ${\bf V}^\dagger$ is the conjugate transpose of $\bf V$. The decomposition can also be written as
\beq
{\bf M}=\sum_{i=1}^r\sigma_i{\bf u}_i{\bf v}_i^\dagger,
\eeq
where $r$ is the rank of $\bf M$, $\sigma_i$ is the $i$-th largest singular value, and ${\bf u}_i$ and ${\bf v}_i$ are the corresponding left and right singular vectors respectively.

Denote $[m] \equiv \{1, 2, \cdots, m\}$. An $m$th-order tensor $\A=(a_{i_1\cdots i_m})$ is a multi-array of $\Pi_{j=1}^m I_j$ entries, where $i_j\in[I_j]$ for $j\in[m]$. $(I_1,I_2,\ldots,I_m)$ is the dimension of $\A$. When $I_1 =I_2= \cdots = I_m=n$, $\A$ is called an $m$th-order $n$-dimensional tensor.

Given a tensor $\A\in\mathbb{C}^{I_1\times I_2\times\cdots\times I_m}$ and a matrix ${\bf{B}}\in\mathbb{C}^{J_k\times I_k}$, their $k$-mode tensor-matrix multiplication
\begin{align}
&(\A\times_k {\bf B})_{i_1i_2\cdots i_{k-1}j_ki_{k+1}\cdots i_m} \nonumber \\
=&\sum_{i_k=1}^{I_k}a_{i_1i_2\cdots i_{k-1}i_ki_{k+1}\cdots i_m}\,b_{j_ki_k},
\end{align}
produces an $I_1\times I_2\times\cdots\times I_{k-1}\times J_k\times I_{k+1}\times\cdots\times I_m$ tensor. The inner product of two tensors $\A,\B\in\mathbb{C}^{I_1\times I_2\times\cdots\times I_m}$, denoted as $\A\cdot\B$, is defined as
\beq
\A\cdot\B=\sum_{i_1=1}^{I_1}\sum_{i_2=1}^{I_2}\cdots\sum_{i_m=1}^{I_m}a^*_{i_1i_2\cdots i_m}b_{i_1i_2\cdots i_m}.
\eeq
Similar to the matrix case, the induced norm $\sqrt{\A\cdot\A}$ is called the Frobenius norm of $\A$, denoted as $\|\A\|_F$. The $l_1$-norm of the tensor $\A$ is defined as $\|\A\|_1=\sum_{i_1=1}^{I_1}\sum_{i_2=1}^{I_2}\cdots\sum_{i_m=1}^{I_m}|a_{i_1i_2\cdots i_m}|$.

For tensor $\A\in\mathbb{C}^{I_1\times I_2\times\cdots\times I_m}$, if there exist matrices ${\bf X}^{(k)}= [\x_1^{(k)}\x_2^{(k)}\cdots\x_{I_k}^{(k)}] \in\mathbb{C}^{I_k\times I_k}$ with $\|\x^{(k)}_{i_k}\|=1$ for $k\in[m]$ and $i_k\in[I_k]$ such that
\beq\label{Tucker}
\A=\BS\times_1{\bf X}^{(1)}\times_2{\bf X}^{(2)}\times_3\cdots\times_m {\bf X}^{(m)},
\eeq
then (\ref{Tucker}) is said to be a Tucker decomposition  of $\A$, and $\BS=(s_{i_1i_2\cdots i_m})$ is called the core tensor of $\A$. Higher order singular value decomposition is a specific orthogonal Tucker decomposition. Specifically, for $\A\in\mathbb{C}^{I_1\times I_2\times\cdots\times I_m}$, the HOSVD \cite{Lie2000} is defined as
\beq\label{HOSVD_form}
\A=\BS\times_1\U^{(1)}\times_2\U^{(2)}\times_3\cdots\times_m\U^{(m)},
\eeq
where the $k$-mode singular matrix $\U^{(k)}=\left[{\bf u}_1^{(k)}{\bf u}_2^{(k)}\cdots {\bf u}_{I_k}^{(k)}\right]$ is a complex unitary $I_k\times I_k$ matrix, the core tensor $\BS\in\mathbb{C}^{I_1\times I_2\times\cdots\times I_m}$ and its subtensors $\BS_{i_k=\alpha}$, of which the $k$th index is fixed to $\alpha\in [I_k]$, have the following properties.


(i) all-orthogonality:

Two subtensors $\BS_{i_k=\alpha}$ and $\BS_{i_k=\beta}$ are orthogonal for $k=1,2,\cdots,m$:
\beqn
\BS_{i_k=\alpha}\cdot\BS_{i_k=\beta}=0\qquad{\rm when}\quad \alpha\neq\beta,
\eeqn

\begin{figure}
  \centering
  \includegraphics[width=8cm]{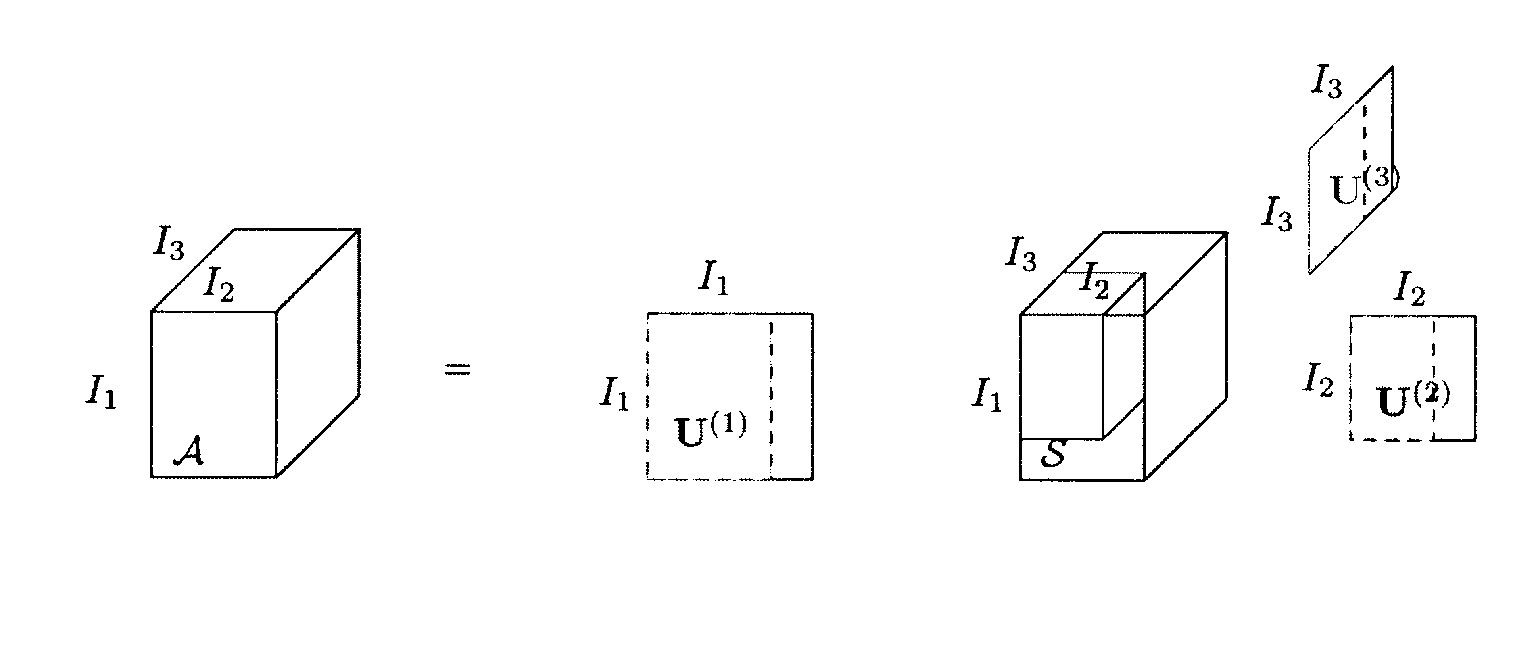}
  \caption{Block diagram of the HOSVD for a third-order tensor. The full lines indicate the full HOSVD in (\ref{HOSVD_form}). The dashed lines and the small block in $\BS$ indicate the truncated HOSVD.}\label{third-order HOSVD}
\end{figure}

(ii) ordering:

Similar to the matrix case, the tensor singular values are defined as the Frobenius norms of the $(m-1)$th-order subtensors of the core tensor $\BS$:
\beqn
\sigma_\alpha^{(k)}=\|\BS_{i_k=\alpha}\|_F,
\eeqn
for $k\in[m]$ and $\alpha \in [I_k]$. Furthermore, these tensor singular values have the following ordering property
\beqn
\sigma_1^{(k)}\geq\sigma_2^{(k)}\geq\cdots\geq\sigma_{I_k}^{(k)}\geq0
\eeqn
for $k\in[m]$.
The block diagram of the HOSVD for a third-order tensor $\A\in\mathbb{C}^{I_1\times I_2\times I_3}$ is described in Fig. \ref{third-order HOSVD}. When $m=2$, i.e., $\A$ is a matrix, the HOSVD is degenerated to the well-known matrix SVD.

HOSVD performs orthogonal coordinate transformations for a higher-order tensor. Here, the unitary matrix $\U^{(k)}$ is also called the $k$-mode factor matrix and considered as the principal components in $k$th mode. Moreover, the entries of the core tensor $\BS$ show the level of interaction among different components.

\begin{figure}
  \centering
  \includegraphics[width=7cm]{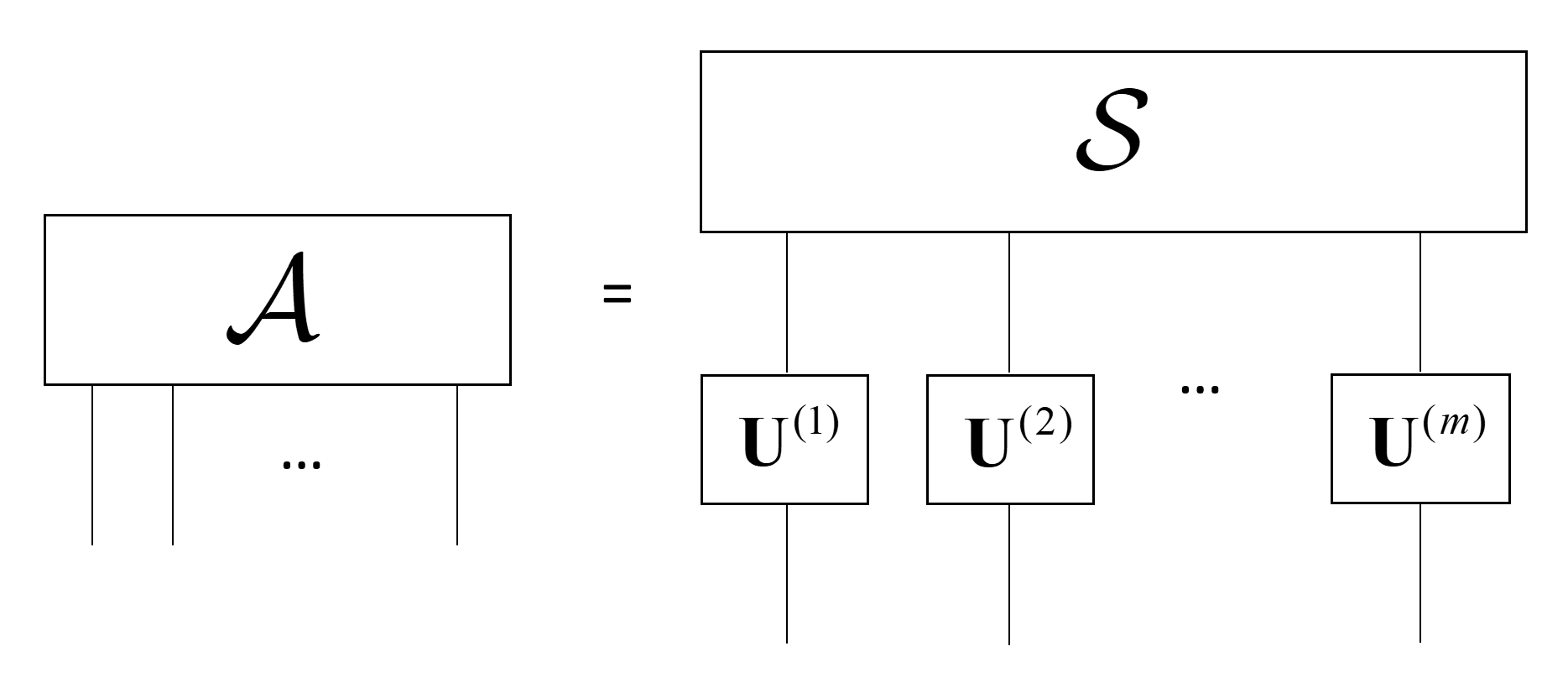}
  \caption{The tensor network notation of equation (\ref{HOSVD_form}).}\label{TN_HOSVD}
\end{figure}

The tensor network notation of HOSVD is depicted in Fig. \ref{TN_HOSVD}. It is known that a tensor corresponds to a multipartite quantum state. Any local unitary transformation on the original tensor can be considered as the local unitary transformation on the corresponding singular matrices, which is vital and useful in quantum computation. The core tensor and singular matrices can also be considered as the two layers in the neural network, with local operations in the first layer and global operations in the second layer.

For an $m$th-order tensor $\A\in\mathbb{C}^{I_1\times I_2\times\ldots\times I_m}$, the mode-$k$ matrix unfolding
$\BA^{(k)}\in\mathbb{C}^{I_k\times(\Pi_{j\neq k}I_j)}$ contains the element $a_{i_1\cdots i_m}$ at the position with row number $i_k$ and column number
\begin{align*}
   & (i_{k+1} - 1)I_{k+2}I_{k+3}\cdots I_m I_1I_2 \cdots I_{k-1}\\
  + & (i_{k+2} - 1)I_{k+3}I_{k+4}\cdots I_m I_1I_2\cdots I_{k-1} + \cdots \\
  + & (i_m - 1)I_1I_2\cdots I_{k-1} + (i_1 - 1)I_2I_3 \cdots I_{k-1}\\
  + & (i_2 - 1)I_3I_4\cdots I_{k-1} + \cdots + i_{k-1}.
\end{align*}
By the above construction, the rank of $\BA^{(k)}$ is at most $I_k$. Clearly, the elements of tensor $\A$ and unfolding matrix $\BA^{(k)}$ have a one-to-one correspondence to each other.

In HOSVD, the columns of $\U^{(k)}$ have been sorted such that the $j$th column ${\bf u}_j^{(k)}$ corresponds to the $j$th largest nonzero singular value of $\BA^{(k)}$. Then, we can similarly define the truncated (or compact) HOSVD \cite{nrhosvd}. For $k\in[m]$, we remain the first $r_k$ columns of $\U^{(k)}$, then $\U^{(k)}\in\mathbb{C}^{I_k\times r_k}$. Finally, the core tensor $\BS$ is of size $r_1\times r_2\times\cdots\times r_m$, and the tuple of numbers $(r_1, r_2, \cdots, r_m)$ is called a multilinear rank. The block diagram of the truncated HOSVD for a third-order tensor is depicted in Fig. \ref{third-order HOSVD}. This truncation is widely used in big data problems. Since the data may be sparse or low-rank, we can take the value of $r_k$ such that $r_k\ll I_k.$ Denote $r=\max_{k \in[m]}r_k$, and $I=\max_{k\in[m]}I_k$. The total number of entries reduces from $I^m$ to $r^m+mIr$.


\section{Q-HOSVD Algorithm 1}\label{a1}
In this section, we present our first Q-HOSVD algorithm.

\begin{algorithm}[H]
\caption{Quantum Higher Order Singular Value Decomposition (Q-HOSVD)}
\label{QHOSVD}
\textbf{Input}:$\,\,\A\in\mbb{C}^{I_1\times\cdots\times I_m},\epsilon,\ket{b}$\\
\textbf{Output}:$\,\,\BS,\U^{(1)},\U^{(2)},\ldots,\U^{(m)}$
\begin{algorithmic}[0]
\STATE 1. Load $\A$ into qRAM.
\FOR{$k=1, \ldots, m$}
%
\STATE 2. Implement qPCA (\ref{qpca}) by the SWAP operator ${\bf S}_{\tilde{A}}^{(k)}$ in  (\ref{SWAP1}).

\STATE 3. Initialize $\ket{0}\ket{\vec 1}\ket{b}$, then apply phase estimation to obtain
\beq\label{PE1}
\ket{\psi}=\sum_{j=0}^{r-1}\beta_j \ket{\tilde{\lambda}_j/N}\ket{\tilde{u}_j}.
\eeq

\STATE 4. Perform measurement on $\ket{\tilde{\lambda}_j/N}$ and extract $\ket{\tilde{u}_j}$ to compose $\U^{(k)}$.
\ENDFOR
\STATE 5. $\BS\leftarrow\A\times_1\U^{(1)^\dagger}\times_2\U^{(2)^\dagger}\times_3\cdots\times_m\U^{(m)^\dagger}.$
\end{algorithmic}
\end{algorithm}


Several techniques and subroutines are applied in Algorithm \ref{QHOSVD}. First, tensor $\A$ to be decomposed is loaded into the quantum register by qRAM. For a fixed $k$, we design a SWAP operator ${\bf S}_{\tilde{A}}^{(k)}$ based on matrix unfolding and Hermitian extension, and harness it to apply qPCA. After that, we initialize the state $\ket{0}\ket{\vec 1}\ket{b}$, where $\ket{b}$ could be any state and considered as a superposition of eigenstates of $\tilde{\BA}^{(k)}$, then apply phase estimation on it to obtain the state $\ket{\psi}$ which is a superposition state composed of eigenvalues and eigenstates. Then, if we hope to obtain the core tensor $\BS$, quantum measurement is performed to reconstruct the singular matrices. Finally, $\BS$ is calculated by the quantum tensor-matrix multiplication among tensor $\A$ and the singular matrices.

%
%
%
%
%
%
%

In the following subsections, we explain the implementation of Algorithm \ref{QHOSVD} step by step. Without loss of generality, we assume $\|\A\|_F=1$, and $I_1=I_2=\cdots=I_m=n$.

\subsection{Step 1}

A vector $\x=(x_1,x_2,\ldots,x_n)^T\in\mathbb{R}^n$ with unit norm can be loaded into a quantum register by an oracle named quantum random access memory (qRAM) \cite{QRAM}:
\beq
\x\mapsto\ket{x}=\sum_{i=0}^{n-1}x_{i+1}\ket{i}
\eeq
with preparation time $O(\log n)$ by a unitary operator $\U_\x$. Note that in quantum computing the indices usually count from 0.

Similarly, a tensor $\A=(a_{i_1i_2\cdots i_m})$ can be accessed by the following multipartite state
\beq
\ket{\Psi}=\sum_{i_1,i_2,\ldots,i_m=0}^{n-1}a_{i_1i_2\cdots i_m}\ket{i_1i_2\cdots i_m},
\eeq
where $i_k=0,\ldots,n-1$ for $k\in[m]$. This procedure can be achieved using $O(m\log n)$ operations by $\U_\A$.

A tree data structure is presented to implement the operation of qRAM \cite{Pra,QRS}, which is a classical data structure with quantum access. In this paper, we extend this data structure for real matrices to both real and complex $m$th-order tensors and summarize it in the following theorem. \\
\bthm\label{data}
For a tensor $\A\in\mathbb{R}(\mathbb{C})^{I_1\times I_2\times\cdots\times I_m}$, there exists a data structure for storing this tensor with quantum access in time $O({\rm polylog}(I_1I_2\cdots I_m))$.
\ethm
\begin{proof}
We prepare a series of unitary operators and append an ancilla $\ket{0}$ at the front when applying every operator
\begin{align}
\U_{m}:&\ket{0}\rightarrow\ket{\phi_m}= \ff{1}{\|\A\|_F}\sum_{i_m=0}^{I_m-1}\|\A(:,\ldots,:,i_m)\|_F\ket{i_m} \nonumber \\
\U_{m-1}:&\ket{0}\ket{\phi_m}\rightarrow\ket{\phi_{m-1}} \nonumber \\
=&\ff{1}{\|\A\|_F}\sum_{i_{m-1}=0}^{I_{m-1}-1}\sum_{i_m=0}^{I_m-1}\ff{\|\A(:,\ldots,:,i_{m-1},i_m)\|_F}{\|\A(:,\ldots,:,i_m)\|_F} \nonumber \\
&\|\A(:,\ldots,:,i_m)\|_F\ket{i_{m-1}}\ket{i_m} \nonumber \\
=&\ff{1}{\|\A\|_F}\sum_{i_{m-1}=0}^{I_{m-1}-1}\sum_{i_m=0}^{I_m-1}\|\A(:,\ldots,:,i_{m-1},i_m)\|_F \nonumber \\
&\ket{i_{m-1}}\ket{i_m} \nonumber \\
&\vdots \nonumber \\
\U_1:&\ket{0}\ket{\phi_2}\rightarrow\ket{\phi_1} \nonumber \\
=&\ff{1}{\|\A\|_F}\sum_{i_1=0}^{I_1-1}\cdots\sum_{i_m=0}^{I_m-1}a_{i_1\cdots i_m}\ket{i_1}\cdots\ket{i_m}.
\end{align}
Basically, the data structure consists of several binary trees named as $B_i^{(k)}$, $i=0,\ldots,I_k-1, k \in [m]$. The top root stores the Frobenius norm of $\A$. The root of each $B_i^{(k)}$ stores the value $\|\A(:,\ldots,:,i_{k+1},\ldots,i_m)\|_F^2$ for $k\in[m-1]$ and the weight of an interior node is just the sum of the weights of its children. The leaf nodes at the bottom store the weights $\A(i_1,\cdots, i_m)^2$ and its sign ${\rm sgn}(\A(i_1,\cdots, i_m))$ for real tensors. For complex tensors, there is one more layer of binary trees for storing the real and imaginary part of a complex entry $\A(i_1,\cdots, i_m)$ and their signs. Therefore, one more qubit is required for storing a complex tensor, and the extra time is $O(1)$ so we can omit it. If a quantum algorithm has access to this data structure, a series of controlled rotations are applied on the initial state in time $O(\lceil\log I_1\rceil\cdots\lceil\log I_m\rceil)$. We consider it as $O({\rm polylog}(I_1I_2\cdots I_m))$.
\end{proof}

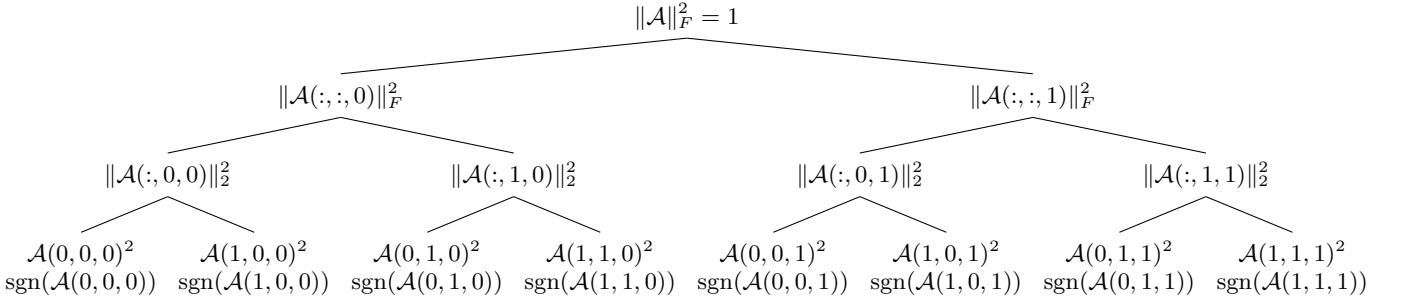
\begin{figure*}
\centering
\small{
\begin{tikzpicture}
\tikzset{every tree node/.style={align=center,anchor=north}}
\Tree [.$\|\A\|^2_F=1$
[.$\|\A(:,:,0)\|^2_F$ [.$\|\A(:,0,0)\|^2_2$ $\A(0,0,0)^2$\\${\rm sgn}(\A(0,0,0))$ $\A(1,0,0)^2$\\${\rm sgn}(\A(1,0,0))$ ]
[.$\|\A(:,1,0)\|^2_2$ $\A(0,1,0)^2$\\${\rm sgn}(\A(0,1,0))$ $\A(1,1,0)^2$\\${\rm sgn}(\A(1,1,0))$ ] ]
[.$\|\A(:,:,1)\|^2_F$ [.$\|\A(:,0,1)\|^2_2$ $\A(0,0,1)^2$\\${\rm sgn}(\A(0,0,1))$ $\A(1,0,1)^2$\\${\rm sgn}(\A(1,0,1))$ ]
[.$\|\A(:,1,1)\|^2_2$ $\A(0,1,1)^2$\\${\rm sgn}(\A(0,1,1))$ $\A(1,1,1)^2$\\${\rm sgn}(\A(1,1,1))$ ] ] ]
\end{tikzpicture}
}
\caption{The data structure for a real $2\times2\times2$ tensor.}
\label{rtree222}
\end{figure*}

\begin{figure*}
\centering
\small{
\begin{tikzpicture}
\tikzset{every tree node/.style={align=center,anchor=north}}
\Tree [.$\|\A\|^2_F=1$
[.$\|\A(:,:,0)\|^2_F$ [.$\|\A(:,0,0)\|^2_2$ [.$|\A(0,0,0)|^2$ ${\rm Re}(\A(0,0,0))^2$\\${\rm sgn}({\rm Re}(\A(0,0,0)))$ ${\rm Im}(\A(0,0,0))^2$\\${\rm sgn}({\rm Im}(\A(0,0,0)))$ ] [.$|\A(1,0,0)|^2$ ${\rm Re}(\A(1,0,0))^2$\\${\rm sgn}({\rm Re}(\A(1,0,0)))$ ${\rm Im}(\A(1,0,0))^2$\\${\rm sgn}({\rm Im}(\A(1,0,0)))$ ] ]
$\|\A(:,1,0)\|^2_2$\\$\bf \vdots$  ]
[.$\|\A(:,:,1)\|^2_F$\\$\bf \vdots$ ] ]
\end{tikzpicture}
}
\caption{The data structure for a complex $2\times2\times2$ tensor.}
\label{ctree222}
\end{figure*}
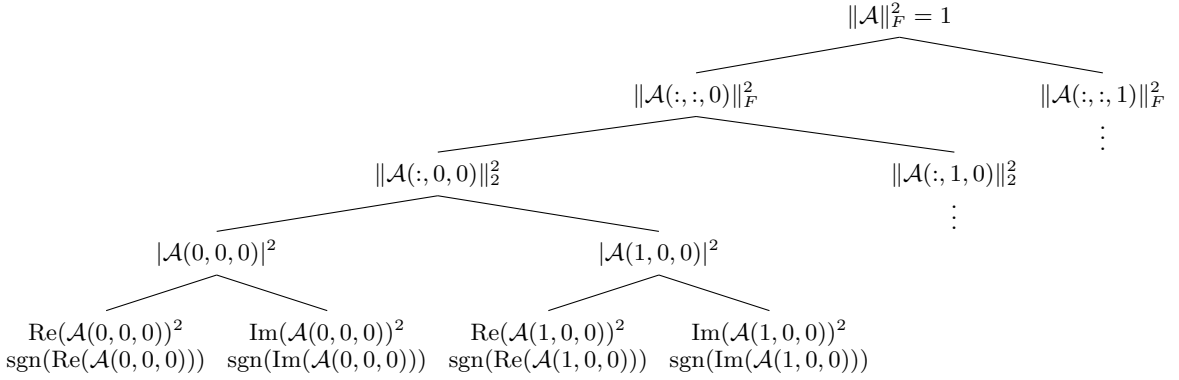

For demonstration, the graph illustrations of the data structure for a real and complex $2\times2\times2$ tensor are given in Fig. \ref{rtree222} and \ref{ctree222} respectively.

\subsection{Step 2}
\begin{figure}
  \centering
  \includegraphics[width=6cm]{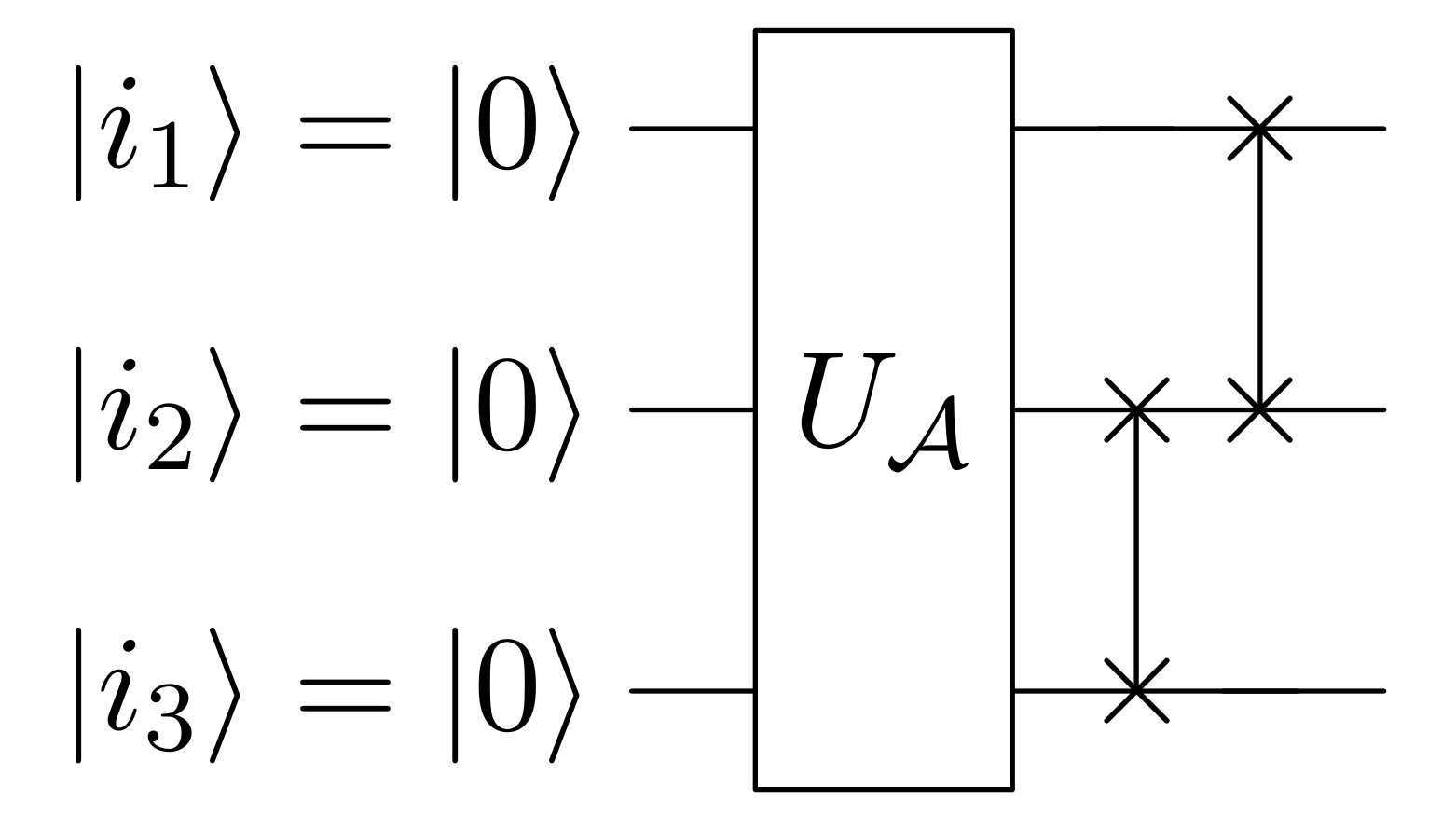}
  \caption{The quantum circuit to perform mode-3 matrix unfolding of a $2\times2\times2$ tensor. The unitary operator $\U_\A$ is the one used in Step 1.}\label{unfold}
\end{figure}

The quantum unfolding matrix $\BA^{(k)}=\left(a^{\prime(k)}_{i_kj_k}\right)$ can be processed by a SWAP operator $\U_{\rm SP}^{(k)}$:
\begin{align}\label{USP}
&\sum_{i_1,i_2,\ldots,i_m=0}^{n-1}a_{i_1i_2\cdots i_m}\ket{i_1i_2\cdots i_m} \nonumber\\
\stackrel{\U_{\rm SP}^{(k)}}{\longrightarrow}&\sum_{i_1,i_2,\ldots,i_m=0}^{n-1} a_{i_1i_2\cdots i_m} \ket{i_ki_{k+1}\cdots i_m i_1\cdots i_{k-1}} \nonumber\\
=&\sum_{i_k=0}^{n-1}\sum_{j_k=0}^{n^{m-1}-1}a^{\prime(k)}_{i_kj_k}\ket{i_kj_k},
\end{align}
where $\ket{j_k}=\ket{i_{k+1}\cdots i_m i_1\cdots i_{k-1}}$. For example, for a $2\times2\times2$ tensor $\A$, the entries correspond to those of mode-3 unfolding matrix $\BA^{(3)}$ by
\begin{align}
a_{000}\ket{000}&\rightarrow a^{\prime(3)}_{00}\ket{00} \nonumber\\
a_{010}\ket{010}&\rightarrow a^{\prime(3)}_{01}\ket{01} \nonumber\\
&\,\,\,\,\vdots \\
a_{101}\ket{101}&\rightarrow a^{\prime(3)}_{12}\ket{12} \nonumber\\
a_{111}\ket{111}&\rightarrow a^{\prime(3)}_{13}\ket{13}. \nonumber
\end{align}
The corresponding SWAP operator is $\U_{\rm SP}^{(3)}=({\rm SWAP}\otimes {\bf I})({\bf I}\otimes{\rm SWAP})$, where $\bf I$ is a $2\times2$ identity matrix, and $\rm SWAP$ is the well-known $4\times4$ SWAP operator ${\rm SWAP}=\sum_{j, \ell=0}^1 |\ell\rangle\langle j|\otimes| j\rangle\langle \ell|$. The circuit of the above operations and tensor input is given in Fig. \ref{unfold}. Except for mode-1 unfolding which does not require SWAP operations, other mode-$k$ unfoldings require $m-1$ SWAP operations. Combined with the complexity of input in Step 1, the total complexity is still $O(m\log n)$.



Denote $N=n+n^{m-1}$. Since $\BA^{(k)}=\left(a^{\prime(k)}_{i j}\right)\in\mathbb{C}^{n\times n^{m-1}}$ is not a Hermitian matrix, we consider the following extended matrix
\begin{align}\label{HE}
\tilde{\BA}^{(k)}:=&\begin{bmatrix} 0 & \BA^{(k)}\\\BA^{(k)^\dagger} & 0\end{bmatrix} \nonumber\\
=&\sum_{i_k=0}^{n-1}\sum_{j_k=0}^{n^{m-1}-1}a^{\prime(k)}_{i_k j_k}\ket{i_k}\bra{j_k+n}+\overline{a^{\prime(k)}_{i_k j_k}} \ket{j_k+n}\bra{i_k} \nonumber \\
=&\sum_{i_1,i_2,\ldots,i_m=0}^{n-1} a_{i_1i_2\cdots i_m} \ket{i_k}\bra{i_{k+1}\cdots i_m i_1\cdots i_{k-1}+n} \nonumber \\
&+\overline{a_{i_1i_2\ldots i_m}}\ket{i_{k+1}\cdots i_m i_1\cdots i_{k-1}+n}\bra{i_k},
\end{align}
where $\ket{i_k}\in\mathbb{C}^{N}$ is the $i_k$-th computational basis. Note that $i_k$ runs from 0 to $n-1$. Then $\tilde{\BA}^{(k)}$ is an $N\times N$ Hermitian matrix. For Hermitian matrices, the singular values are the absolute value of eigenvalues, so phase estimation \cite{Bible} can be used to apply the singular value decomposition. Let $r={\rm rank}(\tilde{\BA}^{(k)})$. Since ${\rm rank}(\BA^{(k)})\leq n$, $r\leq 2n$.

For the Hermitian matrix $\tilde{\BA}^{(k)}$, we define a SWAP-like operator ${\bf S}^{(k)}_{\tilde{A}}\in\mathbb{C}^{N^2\times N^2}$ based on the entries of $\tilde{\BA}^{(k)}$:
\begin{align}\label{SWAP1}
{\bf S}_{\tilde{A}}^{(k)}:=\sum_{\ell,j=0}^{N-1}&\tilde{\BA}_{\ell j}^{(k)}\ket{j}\bra{\ell}\otimes\ket{\ell}\bra{j} \nonumber\\
=\sum_{i_k=0}^{n-1}\sum_{j_k=0}^{n^{m-1}-1}&a^{\prime(k)}_{i_k j_k}\ket{j_k+n}\bra{i_k}\otimes\ket{i_k}\bra{j_k+n} \nonumber\\
+\,&\overline{a^{\prime(k)}_{i_k j_k}} \ket{j_k}\bra{j_k+n}\otimes\ket{j_k+n}\bra{i_k} \nonumber\\
=\sum_{i_1,i_2,\ldots,i_m=0}^{n-1} & a_{i_1i_2\ldots i_m}\ket{i_{k+1}\cdots i_mi_1\cdots i_{k-1}+n}\bra{i_k} \nonumber \\
& \otimes\ket{i_k}\bra{i_{k+1}\cdots i_mi_1\cdots i_{k-1}+n} \nonumber \\
+\,&\overline{a_{i_1i_2\ldots i_m}} \ket{i_k}\bra{i_{k+1}\cdots i_mi_1\cdots i_{k-1}+n} \nonumber \\
&\otimes \ket{i_{k+1}\cdots i_mi_1\cdots i_{k-1}+n}\bra{i_k}.
\end{align}
This operator is one-sparse in a quadratically bigger space, i.e., there is no more than one non-zero entry in every row and column, and its entries are efficiently computable. Therefore, the matrix exponentiation $e^{-\ii {\bf S}_{\tilde{A}}^{(k)}\Delta t}$ is efficiently implemented \cite{ESSH}.

The SWAP-like operator (\ref{SWAP1}) combines the mode-$k$ unfolding (\ref{USP}) and Hermitian extension (\ref{HE}), since they are all related to SWAP operations. It only requires access to the entries of original tensor $\A$.

For simplicity, in the following we use $\BA$ to represent ${\BA}^{(k)}$ when $k$ is fixed.  Let $\rho_1$ and $\rho_2$ be two distinct density matrices, where $\rho_1=\ket{\vec 1}\bra{\vec 1}$ with $\ket{\vec 1}=\frac{1}{\sqrt{N}}\sum_{\ell=0}^{N-1}\ket{\ell}$.

\blem \label{lemmaqpca}
\cite{QPCA} By quantum principal component analysis (qPCA), the unitary $e^{-\ii\frac{\tilde{\BA}}{N}\Delta t}$ is simulated using ${\bf S}_{\tilde{A}}$ through
\elem
\beq\label{qpca}
{\rm tr}_1\{e^{-\ii {\bf S}_{\tilde{A}}\Delta t}\rho_1\otimes\rho_2 e^{\ii {\bf S}_{\tilde{A}}\Delta t}\}\approx e^{-\ii \frac{\tilde{\BA}}{N}\Delta t}\rho_2 e^{\ii \frac{\tilde{\BA}}{N}\Delta t}.
\eeq

Let $\epsilon_0$ be the trace norm of the error term $O(\Delta t^2)$ in (\ref{qpca}). For $s$ steps, the resulting error is $\epsilon_1=s\epsilon_0\leq 2s\|\A\|^2_{\rm max} \Delta t^2$, where $\|\A\|_{\rm max}=\max_{i_1,\ldots,i_m}|a_{i_1\cdots i_m}|$. The simulated time is $t=s\Delta t$. Then,
\beq
\frac{\epsilon_1}{s}\leq2\|\A\|^2_{\rm max} \left(\frac{t}{s}\right)^2.
\eeq
Thus,
\beq\label{step}
s=O\left(\frac{t^2}{\epsilon_1}\|\A\|^2_{\rm max}\right)
\eeq
steps are required to simulate $e^{-\ii \frac{\tilde{\BA}}{N}\Delta t}$ if $\epsilon_1$ and $t$ are fixed.

Since we have assumed $\|\A\|_F=1$, then $\|\A\|_{\rm max}=O(1)$. Hence, $s=O(t^2/\epsilon_1)$. 
Applying the output in equation (\ref{qpca}) again in the second register, we obtain
\begin{align}
&{\rm tr}_1\left\{e^{-\ii {\bf S}_{\tilde{A}}\Delta t}\rho_1\otimes\left(\rho_2-\ii\frac{\Delta t}{N}[\tilde{\BA},\rho_2]+O(\Delta t^2)\right) e^{\ii {\bf S}_{\tilde{A}}\Delta t}\right\}\nonumber\\
=&{\rm tr}_1\{e^{-\ii {\bf S}_{\tilde{A}}\Delta t}\rho_1\otimes\rho_2 e^{\ii {\bf S}_{\tilde{A}}\Delta t}\}\nonumber\\
&-\ii\frac{\Delta t}{N}{\rm tr}_1\{e^{-\ii {\bf S}_{\tilde{A}}\Delta t}(\rho_1\otimes[\tilde{\BA},\rho_2]) e^{\ii {\bf S}_{\tilde{A}}\Delta t}\}+O(\Delta t^2)\nonumber\\
=&\rho_2-\ii\frac{\Delta t}{N}[\tilde{\BA},\rho_2]-\ii\frac{\Delta t}{N}{\rm tr}_1\{\rho_1\otimes[\tilde{\BA},\rho_2]\}+O(\Delta t^2)\nonumber\\
=&\rho_2-\ii\frac{2\Delta t}{N}[\tilde{\BA},\rho_2]+O(\Delta t^2).
\end{align}
Thus, by continuously using $k$ copies of $\rho_1$ we can simulate $e^{-\ii(\tilde{\BA}/N)k\Delta t}$.



\subsection{Step 3}
Next, we use the quantum phase estimation algorithm to estimate the eigenvalues of $e^{-\ii(\tilde{\BA}/N)\Delta t}$.

\blem \cite{Kitaev} \label{lemmape}
Let $\U$ be an $n\times n$ unitary operator, with eigenvectors $\left|v_{j}\right\rangle$ and eigenvalues $e^{\ii \theta_{j}}$ for $\theta_{j} \in[-\pi, \pi],$ i.e. we have $\U\left|v_{j}\right\rangle= e^{\ii \theta_{j}}\left|v_{j}\right\rangle$ for $j =0,1,\ldots,n-1.$ For a precision
parameter $\epsilon>0,$ there exists a quantum phase estimation algorithm that runs in time $O(T(\U) \log n / \epsilon)$ and with probability $1-1 / {\rm poly} (n)$ maps a state $\sum_{j =0}^{n-1} \alpha_{j}\left|v_{j}\right\rangle$ to the state $\sum_{j=0}^{n-1} \alpha_{j}\left|v_{j}\right\rangle\left|\bar{\theta}_{j}\right\rangle$
such that $\bar{\theta}_{j} \in \theta_{j} \pm \epsilon$ for all $j =0,1,\ldots,n-1.$
\elem

\bthm
For the input $\ket{0}^{\otimes d}\ket{\vec 1}\ket{b}$, apply (\ref{qpca}) in Lemma \ref{lemmaqpca} to the input, and apply phase estimation, then the superposition state (\ref{PE1}), where $\ket{\tilde{\lambda}_j/N}$ is the estimated eigenvalue of $\tilde{\BA}/N$ encoded in basis qubits.
\ethm

\begin{proof}
Given an initial quantum state
\beq\label{ini}
\ket{0}^{\otimes d}\ket{\vec 1}\ket{b}=\ket{\underbrace{00\cdots0}_d}\ket{\vec 1}\ket{b}
\eeq
with $d=O(\lceil\log(1/\epsilon_2)\rceil)$ control qubits, where $\ket{b}$ is the superposition of eigenvectors $\ket{\tilde{u}_j}$ corresponding to $\tilde{\lambda}_j$:
\beq\label{inib}
\ket{b}=\sum_{j=0}^{N-1}\beta_j\ket{\tilde{u}_j}, \qquad \sum_{j=0}^{N-1}|\beta_j|^2=1,
\eeq
$\epsilon_2$ is the accuracy for approximating the eigenvalues. Let $\rho_2=\ket{b}\bra{b}$. We first apply Hadamard operators to the first register, then the state (\ref{ini}) becomes
\beq\label{Hada}
\frac{1}{\sqrt{2^d}}\sum_{\ell=0}^{2^d-1} \ket{\ell}\ket{\vec 1}\ket{b},
\eeq
whose density matrix has the following form
\beq\label{dm}
\frac{1}{2^d}\sum_{\ell=0}^{2^d-1}\ket{\ell}\bra{\ell}\otimes\rho_1\otimes\rho_2.
\eeq
Then we multiply $\sum_{\ell=0}^{2^d-1}\ket{\ell}\bra{\ell}\otimes(e^{-\ii {\bf S}_A\Delta t})^\ell$ and $\sum_{\ell=0}^{2^d-1}\ket{\ell}\bra{\ell}\otimes(e^{\ii {\bf S}_A\Delta t})^\ell$ to both sides of (\ref{dm}) to obtain
\beq
\sum_{\ell=0}^{2^d-1}\ket{\ell}\bra{\ell}\otimes\left((e^{-\ii {\bf S}_A\Delta t})^\ell\rho_1\otimes\rho_2(e^{\ii {\bf S}_A\Delta t})^\ell\right).
\eeq
Next, we perform a partial trace to the second register using (\ref{qpca}) resulting in
\beq
\sum_{\ell=0}^{2^d-1}\ket{\ell}\bra{\ell}\otimes\left((e^{-\ii \frac{\tilde{\BA}}{N}\Delta t})^\ell\rho_2(e^{\ii \frac{\tilde{\BA}}{N}\Delta t})^\ell\right).
\eeq
After that, we apply the phase estimation algorithm in Lemma \ref{lemmape} to obtain the estimated eigenvalues of $\tilde{\BA}/N$, since
\begin{align}
e^{-\ii \frac{\tilde{\BA}}{N}\Delta t}\ket{b}&=\sum_{j=0}^{N-1}\beta_j e^{-\ii \frac{\tilde{\BA}}{N}\Delta t}\ket{\tilde{u}_j}\nonumber\\
&=\sum_{j=0}^{N-1}\beta_j e^{-\ii \lambda_j(\frac{\tilde{\BA}}{N})\Delta t}\ket{\tilde{u}_j}.
\end{align}
At last, we implement the inverse quantum Fourier transform \cite{Bible} and remove the first register, the final state (\ref{PE1})
\beqm
\ket{\psi}=\sum_{j=0}^{r-1}\beta_j \ket{\tilde{\lambda}_j/N}\ket{\tilde{u}_j}
\eeqm
is obtained, where $\ket{\tilde{\lambda}_j/N}$ is the estimated eigenvalue of $\tilde{\BA}/N$ encoded in basis qubits. The corresponding eigenvector $\ket{\tilde{u}_j}$ is proportional to $({\bf u}_j;\pm{\bf v}_j)\in\mathbb{C}^N$, where ${\bf u}_j$ and ${\bf v}_j$ are the left and right singular vectors of $\tilde{\BA}$, and the norm of each subvector ${\bf u}_j$ and ${\bf v}_j$ is $1/\sqrt{2}$, independent of their respective lengths $n$ and $n^{m-1}$.
\end{proof}

\subsection{Step 4}

Since ${\BA}$ is of size $n\times n^{m-1}$, ${\BA}$ has at most $n$ singular values $\{\sigma_j\}$. As a result, $\tilde{\BA}$ has at most $2n$ nonzero eigenvalues $\tilde{\lambda}_j\in\{\pm\sigma_j\}$. Next, we measure the first register of state (\ref{PE1}) in the computational basis $\{\ket{0},\cdots,\ket{2^d-1}\}$, all eigenpairs $\ket{\tilde{\lambda}_j/N}\ket{\tilde{u}_j}$ are obtained with probability $|\beta_j|^2$. Discarding the first register and projecting $\ket{\tilde{u}_j}$ onto the ${\bf u}_j$ part by using projection operators ${\bf P}_u=\sum_{i=0}^{n-1}\ket{i}\bra{i}$ and ${\bf P}_v=\sum_{i=n}^{n^{m-1}+n-1}\ket{i}\bra{i}$ result in $\ket{{\bf u}_j}$ with probability $\l \tilde{u}_j|{\bf u}_j,0\r=\frac{1}{2}$. Then, the singular matrix $\U$ is calculated by
\beq
\U=\sum_{j=1}^n \ket{{\bf u}_j}\bra{j}.
\eeq
Repeating measurements with the initial state $\ket{b}=\ket{0},\ket{1},\cdots,\ket{n-1}$ and applying  amplitude amplification \cite{AA}, we can obtain all the singular vectors in $T_U=O(n^{3/2})$ times with probability close to 1. Thus, the singular matrix $\U^{(k)}$ is reconstructed.

\subsection{Step 5}

After we get all $\U^{(k)}$ for $k=1,2,\ldots,m$, in this step we calculate the core tensor $\BS$:
\beq
\BS=\A\times_1\U^{(1)^\dagger}\times_2\U^{(2)^\dagger}\times_3\cdots\times_m\U^{(m)^\dagger}.
\eeq

Here, the calculation is accelerated by the quantum tensor-matrix multiplication, which is similar to the quantum matrix multiplication algorithm by swap test \cite{QMM}. We may calculate $\A\times_k\U^{(k)^\dagger}$ through the following state
\begin{align}\label{tmm}
\frac{1}{\|\A\|_F\|\U^{(k)}\|_F}\sum_{i_1,\ldots,i_{k-1},j_k,i_{k+1},\ldots,i_m=0}^{n-1}\|\U^{(k)}_{\bullet j_k}\|_2 \nonumber\\
\|\A_{i_1\cdots i_{k-1}\bullet i_{k+1}\cdots i_m}\|_2\l\A_{i_1\cdots i_{k-1}\bullet i_{k+1}\cdots i_m}|\U^{(k)}_{\bullet j_k}\r \nonumber\\
\ket{i_1,\cdots,i_{k-1},j_k,i_{k+1},\cdots,i_m}\ket{0}+\ket{0}^\perp,
\end{align}
where $\ket{\A_{i_1\cdots i_{k-1}\bullet i_{k+1}\cdots i_m}}$ is an $n$-level quantum state ($n$-entry vector) if $i_1,\ldots,i_{k-1},i_{k+1},\ldots,i_m$ are all fixed.

By (\ref{tmm}), the success probability is
\begin{align}
&\sum_{i_1,\ldots,i_{k-1},j_k,i_{k+1},\ldots,i_m=0}^{n-1}\|\U^{(k)}_{\bullet j_k}\|_2^2\|\A_{i_1\cdots i_{k-1}\bullet i_{k+1}\cdots i_m}\|_2^2 \nonumber\\
&\l\A_{i_1\cdots i_{k-1}\bullet i_{k+1}\cdots i_m}|\U^{(k)}_{\bullet j_k}\r^2{\Big /}\left(\|\A\|^2_F\|\U^{(k)}\|^2_F\right)\nonumber\\
=&\frac{\|\A\times_k\U^{(k)^\dagger}\|_F^2}{\|\A\|_F^2\|\U^{(k)}\|_F^2}.
\end{align}
Note that unitary matrices preserve norms, and we have assumed that $\|\A\|_F=1$, therefore
\beq
\|\A\times_k\U^{(k)^\dagger}\|_F=\|\A\|_F=1.
\eeq
Thus, after post-selecting $\ket{0}$, the state (\ref{tmm}) becomes
\begin{align}\label{psik}
\ket{\psi^{(k)}}:=&\sum_{i_1,\ldots,i_{k-1},j_k,i_{k+1},\ldots,i_m=0}^{n-1}\|\U^{(k)}_{\bullet j_k}\|_2 \nonumber\\
&\|\A_{i_1\cdots i_{k-1}\bullet i_{k+1}\cdots i_m}\|_2\l\A_{i_1\cdots i_{k-1}\bullet i_{k+1}\cdots i_m}|\U^{(k)}_{\bullet j_k}\r \nonumber \\ &\ket{i_1,\cdots,i_{k-1},j_k,i_{k+1},\cdots,i_m}.
\end{align}
$\ket{\psi^{(k)}}$ corresponds to the tensor $\A\times_k\U^{(k)^\dagger}$, whose amplitudes are exactly the entries of tensor $\A\times_k\U^{(k)^\dagger}$.

After applying amplitude amplification \cite{AA}, the computational complexity is
\begin{align}
T_M&=\tilde{O}\left(\frac{\|\A\|_F\|\U^{(k)}\|_F} {\epsilon_3\|\A\times_k\U^{(k)^\dagger}\|_F}\right) \nonumber \\
&=\tilde{O}\left(\frac{\|\U^{(k)}\|_F}{\epsilon_3}\right) =\tilde{O}\left(\ff{\sqrt{n}}{\epsilon_3}\right)
\end{align}
to accuracy $\epsilon_3$.

Repeating the multiplication between $\A$ and $\U^{(k)^\dagger}$ for $k=1,2,\ldots,m$, we obtain the final state
\beq
\ket{\psi_f}=\sum_{j_1,\ldots,j_m=0}^{n-1}s_{j_1\cdots j_m}\ket{j_1,\cdots,j_m},
\eeq
corresponding to the core tensor $\BS$. The total complexity is $\tilde{O}\left(\ff{m\sqrt{n}}{\epsilon_3}\right)$.


Without loss of generality, we can regard the accuracy of matrix exponentiation, phase estimation and tensor-matrix multiplication as the same value, i.e. $\epsilon_1=\epsilon_2=\epsilon_3=:\epsilon$.

\section{Q-HOSVD Algorithm 2}\label{a2}
For Algorithm \ref{QHOSVD} to be efficiently implemented, the unfolding matrices are required to be low-rank. This result is analyzed in Section \ref{comp} for complexity analysis. However, in some cases the input may not have such good structure. In this section we propose an alternative quantum HOSVD algorithm which is based on quantum singular value estimation (QSVE) \cite{QRS}. In this method, the input is a general matrix, not required to be sparse, low-rank or well-conditioned.

In \cite{Pra,QRS}, the authors introduce a classical tree data structure with quantum access, stated in Lemma \ref{lemmads}, where the quantum states are efficiently prepared corresponding to the rows and columns of matrices. Based on this data structure, a fast quantum algorithm to perform singular value estimation stated in \ref{lemmaqsve} is designed.

\blem \label{lemmads} \cite{QRS} Consider a matrix $\BA=(a_{ij}) \in \mathbb{R}^{N_1\times N_2}$ with $\omega$ nonzero entries. Let $\BA_i$ be its $i$-th row of $\BA$, and $\hat{\BA}=
\frac{1}{\|\BA\|_F}\left[\|\BA_0\|_2 , \|\BA_1\|_2 , \cdots , \|\BA_{N_1-1}\|_2 \right]^\top.$ There exists a data structure storing the matrix $\BA$ in $O(\omega {\rm log}^2(N_1N_2))$ space such that a quantum algorithm having access to the data structure can perform the mapping
\begin{align}
\U_{P}: \ket{i}\ket{0}\rightarrow& \ket{i}\ket{\BA_i}=\frac{1}{\|\BA_i\|_2}\sum_{j=0}^{N_2-1}a_{ij}\ket{i}\ket{j}
	\nonumber \\
&{\rm for}\quad i=0,\cdots,N_1-1; \nonumber \\
\U_{Q}: \ket{0}\ket{j}\rightarrow& \ket{\hat{\BA}}\ket{j}=\frac{1}{\|\BA\|_F}\sum_{i=0}^{N_1-1}\|\BA_i\|_2\ket{i}\ket{j} \nonumber \\
&{\rm for}\quad j=0,\cdots,N_2-1
\end{align}
in time ${\rm polylog}(N_1N_2)$.
\elem

In simple terms, there exists a unitary operator $\U_A$ that prepares $\BA$ by
\beq
\U_A(\ket{0}^{\log N_1}\ket{0}^{\log N_2})=\frac{1}{\|\BA\|_F}\sum_{i,j}a_{ij}\ket{i}\ket{j}
\eeq
in $O({\rm polylog}(N_1N_2))$ time.


Define two isometries $\MP \in \mathbb{C}^{N_1N_2 \times N_1}$ and $\MQ\in \mathbb{C}^{N_1N_2 \times N_2}$ related to $\U_{P}$ and $\U_{Q}$:
\begin{align} \label{PQ}
\MP=\sum_{i=0}^{N_1-1}\ket{i}\ket{\BA_i}\bra{i} , \quad \MQ=\sum_{j=0}^{N_2-1}\ket{\hat{\BA}}\ket{j}\bra{j}.
\end{align}
It can be proved that $\MP^\dagger \MP=\MI_{N_1}, \MQ^\dagger \MQ=\MI_{N_2}$, $2\MP\MP^{\dagger}-\MI_{N_1N_2}$ is unitary and it can be efficiently implemented in time $O({\rm polylog}(N_1N_2))$ in the form of $\U_P$ and $\U_Q$. Actually,
\begin{align} 2\MP\MP^{\dagger}-I_{N_1N_2}&=2\sum_{i}\ket{i}\ket{\BA_i}\bra{i}\bra{\BA_i}-\MI_{N_1N_2} \nonumber\\
&=\U_{P}{\bf G}_P\U_{P}^{\dagger},
\end{align}
where ${\bf G_P}:=2\sum_{i}\ket{i}\ket{0}\bra{i}\bra{0}-\MI_{N_1N_2}$ is a reflection. Similarly, $2\MQ\MQ^{\dagger}-\MI_{N_1N_2}=\U_Q{\bf G}_Q\U_Q^\dagger$, where ${\bf G}_Q:=2\sum_{j}\ket{0}\ket{j}\bra{0}\bra{j}-\MI_{N_1N_2}$.
	
Now denote
\begin{align}\label{W}
\W &=(2\MP\MP^{\dagger}-\MI_{N_1N_2})(2\MQ\MQ^{\dagger}-\MI_{N_1N_2}) \nonumber \\
&=\U_P{\bf G}_P\U_P^\dagger \U_Q{\bf G}_Q\U_Q^\dagger.
\end{align}
All the factors are unitary operators. Then, $\W$ is used for phase estimation to obtain the singular values. The eigenvalue of $\bf W$ is $e^{\ii\theta_i}$ such that $\cos(\theta_i/2) = \sigma_i/\|\BA\|_F$, where $\sigma_i$ is the singular value of $\BA$. The result of QSVE is summarized in the following lemma:

\blem \label{lemmaqsve} \cite{QRS}
Let $\BA \in \mathbb{R}^{N_1 \times N_2}$ stored in the data structure stated in Lemma \ref{lemmads}, and the singular value decomposition of $\BA$ can be written as $\BA=\sum_{\ell=0}^{r-1}\sigma_{\ell}\ket{u_{\ell}}\bra{v_{\ell}}$, where $r=\min (N_1, N_2)$. Let $\epsilon > 0$ be the precision parameter. Then the quantum singular value estimation runs in $O({\rm polylog}(N_1N_2)/{\epsilon})$ and achieves
$\sum_i\beta_i\ket{v_i}\ket{0}\mapsto \sum_i\beta_i\ket{v_i}\ket{\overline{\sigma}_i},$ and $|\overline{\sigma}_i-\sigma_i| \leq \epsilon\|\BA\|_F$ for all $i$ with probability at least $1-1/{\rm poly}(N_2)$.
\elem

\begin{figure}
  \centering
  \includegraphics[width=8cm]{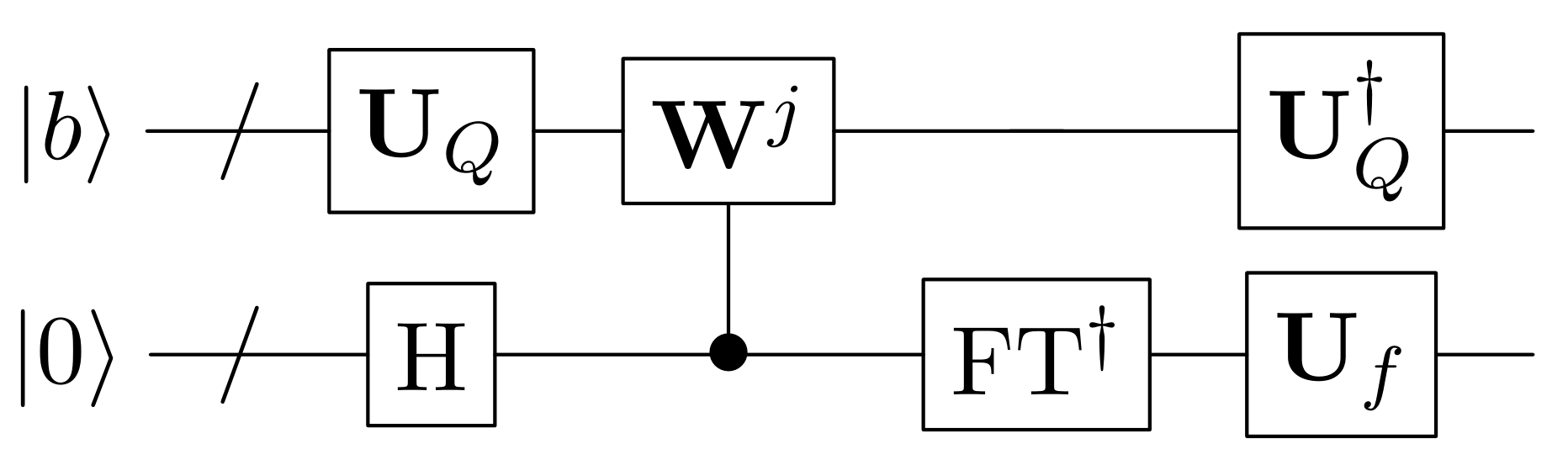}
  \caption{Circuit of QSVE on matrix $\BA$. The second register has $t$ qubits, indicating that the accuracy is $2^{-t}$, and $j$ takes the value of $2^0,2^1,\ldots,2^{t-1}$ respectively. ${\bf U}_f$ maps the singular values of $\bf W$ to those of $\bf A$.}\label{c2}
\end{figure}

The circuit of QSVE on a single matrix is shown in Fig. \ref{c2}. Note that QSVE obtains the superposition state of estimated singular values and the corresponding right singular vectors. Since in HOSVD we aim to obtain the left singular vectors, we can obtain the left singular vectors $\ket{u_i^{(k)}}$ of $\BA^{(k)}$ from the right singular vectors of its conjugate transpose $\BA^{(k)^\dagger}$. Thus, we perform the QSVE on the unfolding matrix $\BA^{(k)^\dagger}$. The Q-HOSVD algorithm based on QSVE is given in Algorithm \ref{QHOSVD2}.

\begin{algorithm}[H]
\caption{Q-HOSVD by QSVE}
\label{QHOSVD2}
\textbf{Input}:$\,\,\A\in\mbb{C}^{I_1\times\cdots\times I_m},\epsilon,\ket{b}$\\
\textbf{Output}:$\,\,\BS,\U^{(1)},\U^{(2)},\ldots,\U^{(m)}$
\begin{algorithmic}[0]
\STATE 1. Prepare the initial state $\frac{1}{\sqrt{m}}\sum_{k=0}^{m-1}\ket{b}\ket{0}\ket{k}.$

\STATE 2. Implement the controlled-$k$ QSVE by the last register to obtain the superposition state
\beqm
\ket{\psi}=\frac{1}{\sqrt{m}}\sum_{k=0}^{m-1}\sum_{i=0}^{r-1}\beta_i^{(k)}\ket{u_i^{(k)}}\ket{\overline{\sigma}_i^{(k)}}\ket{k}.
\eeqm

\STATE 3. Post-select $k$ and perform measurement on $\ket{\overline{\sigma}_i^{(k)}}$ and extract $\ket{u_i^{(k)}}$ to compose $\U^{(k)}$.

\STATE 4. $\BS\leftarrow\A\times_1\U^{(1)^\dagger}\times_2\U^{(2)^\dagger}\times_3\cdots\times_m\U^{(m)^\dagger}$.
\end{algorithmic}
\end{algorithm}

Algorithm \ref{QHOSVD2} is similar to Algorithm \ref{QHOSVD}, but we do not need to apply phase estimation on the extended Hermitian matrices.

For Step 1, we prepare the initial state
\beq
\ket{\psi_0}=\frac{1}{\sqrt{m}}\sum_{k=0}^{m-1}\ket{b}\ket{0}\ket{k},
\eeq
where the first register could be any state and always expressed as a superposition of eigenstates in (\ref{inib}), the second register stores the eigenvalues after phase estimation, and the last register is the index for mode-$k$ unfolding.

For Step 2, assume that the tensor $\A$ is $m$th-order $n$-dimensional for simplicity. Recall that for fixed $k$, $\ket{j_k}=\ket{i_{k+1}\cdots i_m i_1\cdots i_{k-1}}$ as the same in (\ref{USP}), where $j_k=0,\ldots,n^{m-1}-1$. Denote $\ket{\BA_{j_k}}$ be the tube $\ket{\A_{i_1\cdots i_{k-1}\bullet i_{k+1}\cdots i_m}}$. Different from Algorithm \ref{QHOSVD}, we directly prepare the mode-$k$ unfolding matrix through the unitary operators $\U_P^{(k)}$ and $\U_Q^{(k)}$ as in Lemma \ref{lemmads} according to the mode of unfolding:
\begin{align}
\U_{P}^{(k)}: \ket{j_k}\ket{0}\rightarrow& \ket{j_k}\ket{\overline{\BA}_{j_k}}=\frac{1}{\|\BA_{j_k}\|_2}\sum_{i_k=0}^{n-1} \overline{a}_{i_1i_2\cdots i_m} \ket{j_k}\ket{i_k}
	\nonumber \\
&{\rm for}\quad j_k=0,\cdots,n^{m-1}-1; \nonumber \\
\U_{Q}^{(k)}: \ket{0}\ket{i_k}\rightarrow& \ket{\hat{\BA}_k}\ket{i_k}=\frac{1}{\|\A\|_F}\sum_{j_k=0}^{n^{m-1}-1} \|\BA_{j_k}\|_2\ket{j_k}\ket{i_k} \nonumber \\
&{\rm for}\quad i_k=0,\cdots,n-1.
\end{align}
By this way, the above two operators are prepared in time $O(m\,{\rm polylog}n)$, corresponding to $\BA^{(k)^\dagger}$. Then we implement the controlled-$k$ singular value estimation in parallel by applying the operator $\sum_{k=0}^{m-1}{\bf W}^{(k)}\otimes\ket{k}\bra{k}$ on the initial state $\ket{\psi_0}$.  Finally, we undo the phase estimation and apply the inverse of operator $\U_P^{(k)}$, obtaining the state
\beq\label{PE2}
\ket{\psi}=\frac{1}{\sqrt{m}}\sum_{k=0}^{m-1}\sum_{i=0}^{r-1}\beta_i^{(k)}\ket{u_i^{(k)}}\ket{\overline{\sigma}_i^{(k)}}\ket{k}.
\eeq

After that, similar to the Steps 4 and 5 in Algorithm \ref{QHOSVD}, we can make measurements and obtain the singular matrices and core tensor.

\section{Complexity Analysis}\label{comp}
For simplicity, we assume the input is an $m$th-order $n$-dimensional tensor.

For the first Q-HOSVD algorithm, the computational complexity mainly comes from data access, matrix exponential simulation and phase estimation. The data input time is $O(m\,{\rm polylog} n)$. At a simulation time $t$, only the eigenvalues of $\tilde{\BA}^{(k)}/N$ with $|\tilde{\lambda}_j|/N = \Omega(1/t)$ matter \cite{QSVD}, and the eigenvalues smaller than $\epsilon$ are omitted. Note that $\BA^{(k)}$ is an $n\times n^{m-1}$ matrix. For a fixed $k$, let the number of these eigenvalues be $r \leq 2n$, then by
\beq \label{eq:1 feb4}
\left\{
\begin{array}{lcl}
{\rm tr}\{\tilde{\BA}_r^2/N^2\}=\sum_{j=0}^{r-1}\tilde{\lambda}_j^2/N^2=\Omega(r/t^2),\\
{\rm tr}\{\tilde{\BA}_r^2/N^2\}\leq{\rm tr}\{\tilde{\BA}^2/N^2\}=\|\tilde{\BA}\|_F^2/N^2\leq\|\A\|_{\max}^2,
\end{array} \right.
\eeq
we find that the rank of the effectively simulated matrix is $r = O(\|\A\|_{\max}^2t^2)$. By (\ref{step}), there are $O\left(t^2\|\A\|^2_{\rm max}/\epsilon\right)$ steps required to simulate $e^{-\ii \frac{\tilde{\BA}}{N}t}$, where $\|\A\|_{\rm max}=O(1)$, and $1/\epsilon$ can be chosen as $O({\rm polylog}n)$. To make this algorithm efficient, $t=O({\rm polylog}n)$, then the rank $r=O({\rm polylog}n)$, i.e. the matrices have to be low-rank. Thus, the time to implement phase estimation is $O({\rm polylog}n\cdot\log(n)/\epsilon)$. Therefore, the total computational complexity of obtaining (\ref{PE1}) is $O(m^2 {\rm polylog}n)$.

For the second Q-HOSVD algorithm, for each unfolding the time to access the data structure is $O(m\,{\rm polylog}n)$, and the time to implement QSVE is $O(m\,{\rm polylog}n/\epsilon)$, so the total complexity of obtaining (\ref{PE2}) is $O(m^3 {\rm polylog}n)$. Furthermore, there are no requirement for the structure of the input tensor.

Usually, in practical problems, $m\ll n$. Thus, we can omit the order $m$, and the algorithm runs polylogarithmically in the dimensions.

If we want to obtain the singular matrices and core tensor explicitly in the quantum register, we need to make measurements on the states (\ref{PE1}) and (\ref{PE2}), and reconstruct singular matrices and finally calculate the core tensor by quantum tensor matrix multiplication, the complexities are $O(m^3n^{2}{\rm polylog} n)$ and $O(m^4n^{2}{\rm polylog} n)$ respectively.

\section{Application on Recommendation Systems}\label{rs}
%
In \cite{Tagre}, the authors make use of HOSVD for tag recommendations. Given an initial third-order tensor with usage data triplets (user, item, tag), they implement HOSVD and do truncations to obtain the core tensor and reconstructed tensor with smaller dimensions. Then, based on the entries of the reconstructed tensor, the tags are recommended to users. We have carried out the similar SVD and truncation operations in \cite{tsvd} by another tensor decomposition called t-svd.

In this section, we introduce a {\it hybrid} quantum-classical recommendation method for context-aware collaborative filtering (CF) based on tensor factorization (TF), named as multiverse recommendation \cite{Mulre}. TF is an extension of matrix factorization (MF) to multiple dimensions. HOSVD is chosen as our TF approach to analyze the recommendation systems, due to its relevance among the different categories. Given the known preference tensor, we use HOSVD model to find out the missing information. This problem is well-known as the completion problem in recommendation systems \cite{TC}. Our contribution is designing a {\it hybrid} quantum-classical recommendation algorithm to accelerate this process.

Context has been universally acknowledged as an important factor for analyzing recommendation systems. A pair (user, item) is extended to a triplet (user, item, context) or even larger multiplets, where context denotes the factor that may influence a user's preference on a specified item, e.g. time, location, and we consider the interactions among them. Generally, any number of contexts can be added to this recommendation system, and the correlation is described by a relevance score function $f_{rs}$ as follows:
\begin{align}\label{frs}
  f_{rs}: & \, {\rm User}\times {\rm Item}\times {\rm Context}_1\times\cdots \times{\rm Context}_\ell \nonumber \\
   & \longrightarrow {\rm Relevance\,\, Score}
\end{align}
with $\ell$ different contexts, so that the total number of dimensions is $\ell+2$. Thus, we can use a tensor with order $m=\ell+2$ to express the set of relevance scores, and utilize HOSVD model to describe it. Such methods are widely applied in recommendation systems like Netflix prize problems \cite{NF} and so on.

Denote the given preference tensor $\Y\in\{0,\ldots,5\}^{I_1\times I_2\times\cdots\times I_m}$ containing the observed ratings ranging from 1 to 5, and value 0 indicates that the item has not been rated yet. The aim is to find out such missing values and give a good recommendation to users. Denote the factor matrices $\U\in\mathbb{R}^{I_1\times d_1}$, $\M\in\mathbb{R}^{I_2\times d_2}$, $\C^{(k)}\in\mathbb{R}^{I_{k+2}\times d_{k+2}}$, for $k=1,\ldots,\ell$. Then, $\BS\in\mathbb{R}^{d_1\times d_2\times\cdots\times d_m}$. Let ${\bf{d}}=[d_1,d_2,\cdots,d_m]^\top$, and $d=\max_{j\in[m]}d_j$.

To obtain the recommendations based on HOSVD, we design a loss function and optimize over it. The loss function is characterized as $L(\T(\theta),\Y)$, where $\theta$ is the model parameter, i.e., $\theta:=(\BS,\U,\M,\C^{(1)},\C^{(2)},\ldots,\C^{(\ell)})$. Denote a set $D:=\{(i_1,i_2,\ldots,i_m)\,\,|\,\, y_{i_1i_2\cdots i_m}>0\}$ an observation history, and $K:=|D|$ the number of observed ratings. The total loss function is defined as
\beq
L(\mathcal{T}(\theta), \mathcal{Y}):=\frac{1}{\|\BS\|_{1}} \sum_{(i_1,i_2,\ldots,i_m) \in D} l\left( t_{i_1i_2\cdots i_m}, y_{i_1i_2\cdots i_m}\right),
\eeq
which only applies on the observed values in $\Y$. Function $l(t,y)$ is a pointwise loss function, that can be based on $l_2$ norm, e.g., $l(t,y)=(t-y)^2/2$, or other types of distance measure. By adding a regularization term to avoid overfitting, we establish the objective function
\beq
J(\theta):=L(\mathcal{T}(\theta), \mathcal{Y})+\Omega(\theta)
\eeq
with trivial regularizers
\beq
\Omega(\theta)=\lambda_1\|\U\|^2_F+\lambda_2\|\M\|^2_F+\sum_{k=1}^\ell\lambda_{k+2}\|\C^{(k)}\|^2_F +\lambda_\BS\|\BS\|^2_F.
\eeq
Usually, the parameters of matrices can be chosen as the same value, i.e., $\lambda_1=\lambda_2=\cdots=\lambda_m=:\lambda$.

We optimize these matrices and the core tensor by stochastic gradient descent (SGD) method \cite{SGD}. SGD randomly picks a sample at one time and perform gradient descent. It usually compares with batch gradient descent (BGD) which runs over all the samples each iteration. BGD converges globally in every step but it is computationally prohibitive for our problem. The cost of SGD is low, but it usually converges in a local minimum. For big data problems, SGD often converges without running over all the samples. The whole tensor completion algorithm based on the HOSVD model is given in Algorithm \ref{TF}.
\begin{algorithm}[H]
\caption{Tensor Completion by HOSVD}
\label{TF}
{\textbf{Input:}} $\Y,{\bf{d}},\eta,\lambda,\lambda_\BS$\\ 
{\textbf{Output:}} $\BS,\U,\M,\C^{(1)},\cdots,\C^{(\ell)}$\\ 
Initialize $\U,\M,\C^{(1)},\cdots,\C^{(\ell)},\BS$ with small values, $\T$ with all zeros.
\begin{algorithmic}
\WHILE{$(i_1,i_2,\ldots,i_m)\in D$}
\STATE $ t_{i_1i_2\cdots i_m}=\BS \times_{1} \U_{i_1 \bullet} \times_{2} \M_{i_2 \bullet} \times_3 \C_{i_3 \bullet}^{(1)}\times_4\cdots\times_{m} \C^{(\ell)}_{i_m \bullet}$
\STATE $\U_{i_1 \bullet} \longleftarrow \U_{i_1 \bullet}-\eta \lambda \U_{i_1 \bullet}-\eta \partial_{\U_{i_1 \bullet}} l\left( t_{i_1i_2\cdots i_m},  y_{i_1i_2\cdots i_m}\right)$
\STATE $\M_{i_2 \bullet} \longleftarrow \M_{i_2 \bullet}-\eta \lambda \M_{i_2 \bullet}-\eta \partial_{\M_{i_2 \bullet}} l\left( t_{i_1i_2\cdots i_m},  y_{i_1i_2\cdots i_m}\right)$
\STATE $\C^{(1)}_{i_3 \bullet} \longleftarrow \C^{(1)}_{i_3 \bullet}-\eta \lambda \C^{(1)}_{i_3 \bullet}-\eta \partial_{\C^{(1)}_{i_3 \bullet}} l\left( t_{i_1i_2\cdots i_m},  y_{i_1i_2\cdots i_m}\right)$
\STATE $\quad\quad\quad\vdots$
\STATE $\C^{(\ell)}_{i_m \bullet} \longleftarrow \C^{(\ell)}_{i_m \bullet} -\eta \lambda \C^{(\ell)}_{i_m \bullet} -\eta \partial_{\C^{(\ell)}_{i_m \bullet} } l\left( t_{i_1i_2\cdots i_m},  y_{i_1i_2\cdots i_m}\right)$
\STATE $\BS\longleftarrow\BS-\eta\lambda_\BS\BS-\eta\partial_\BS l\left( t_{i_1i_2\cdots i_m},  y_{i_1i_2\cdots i_m}\right)$
\ENDWHILE
\end{algorithmic}
\end{algorithm}

%
%

Algorithm \ref{TF} can be considered as a training method by SGD. After we obtain the factor matrices and core tensor, $\T$ is computed explicitly by
\beq
\T = \BS \times_1 \U \times_2\M \times_3 \C^{(1)} \times_4 \cdots \times_m \C^{(\ell)}
\eeq
as an approximation of the preference tensor $\mathcal{Y}$ and we give recommendations to users according to the entries of $\T$.

%
%
%
%
%
%
%
%
%
%

This algorithm is a {\it hybrid} quantum-classical algorithm. The computation of gradients is accelerated by some quantum subroutines, and the rest procedures are performed by classical computers.
Denote $\circ$ the outer product between tensors. The gradients are, e.g.,
\begin{align}\label{Ui}
&\partial_{\U_{i_1 \bullet}} l\left( t_{i_1i_2\cdots i_m},  y_{i_1i_2\cdots i_m}\right) \nonumber \\
=&\partial_{ t_{i_1i_2\cdots i_m}} l\left( t_{i_1i_2\cdots i_m},  y_{i_1i_2\cdots i_m}\right) \BS \times_2 \M_{i_2 \bullet} \times_3 \cdots \times_{m} \C^{(\ell)}_{i_m \bullet},
\end{align}
\begin{align}
&\partial_{\BS} l\left( t_{i_1i_2\cdots i_m},  y_{i_1i_2\cdots i_m}\right) \nonumber \\
=&\partial_{ t_{i_1i_2\cdots i_m}} l\left( t_{i_1i_2\cdots i_m},  y_{i_1i_2\cdots i_m}\right) \U_{i_1 \bullet} \circ \M_{i_2 \bullet} \circ \cdots \circ \C^{(\ell)}_{i_m \bullet}.
\end{align}

For gradient (\ref{Ui}), $\partial_{ t_{i_1i_2\cdots i_m}} l\left( t_{i_1i_2\cdots i_m},  y_{i_1i_2\cdots i_m}\right)$ is a simple function, if the loss function takes $l_2$ norm, then $\partial_t l(t,y)=t-y$. Define ${\bf g}(\U_{i_1\bullet})=\BS \times_2 \M_{i_2 \bullet} \times_3 \cdots \times_{m} \C^{(\ell)}_{i_m \bullet}.$ The entry ${\bf g}_j(\U_{i_1\bullet})$ is equivalent to
\beq
\BS_{i_1=j}\cdot (\M_{i_2 \bullet} \circ \cdots \circ \C^{(\ell)}_{i_m \bullet})=:\BS_{i_1=j}\cdot{\cal Z},
\eeq
the inner product of two $(m-1)$th order tensors. By quantum matrix multiplication algorithm \cite{QMM}, the outer product of two vectors $\ket{a}$ and $\ket{b}$ can be performed in time $O(\|a\|_F^3\|b\|_F^3{\rm polylog}n/\epsilon_4\|a\circ b\|_F^3)=O({\rm polylog}n/\epsilon_4)$ to accuracy $\epsilon_4$, which holds for classical vectors. Thus, we can perform the quantum outer product $\M_{i_2 \bullet} \circ \cdots \circ \C^{(\ell)}_{i_m \bullet}$ in time $O({\rm polylog}n/\epsilon_4)$. The resulting state is
\beq
\ket{z}=\frac{1}{\|{\cal Z}\|_F}\sum_{i_2,i_3,\ldots,i_m=0}^{n-1}z_{i_2i_3\cdots i_m}\ket{i_2}\ket{i_3}\cdots\ket{i_m}.
\eeq
Next, we load the subtensor $\BS_{i_1=j}$ into the quantum register as
\beq
\ket{s}=\frac{1}{\|{\cal S}_{i_1=j}\|_F}\sum_{i_2,i_3,\ldots,i_m=0}^{n-1}s_{ji_2i_3\cdots i_m}\ket{i_2}\ket{i_3}\cdots\ket{i_m}.
\eeq
Then, we can construct the following superposition state:
\begin{align}\label{}
  \ket{\phi_1} & =\frac{1}{\sqrt{2}}(\ket{+}\ket{s}+\ket{-}\ket{z}) \nonumber\\
   & =\sin\theta\ket{0}\ket{u}+\cos\theta\ket{1}\ket{v}
\end{align}
with $\cos\theta=\sqrt{(1-\l s|z\r)/2}$. Then, by applying quantum amplitude estimation algorithm \cite{QAE}, we can obtain $h$ such that
\beq
\left|\frac{1-\l s|z\r}{2}-h\right|\leq\epsilon_5
\eeq
in time $O(m\log d/\epsilon_5)$. Therefore, $\|\BS_{i_1=j}\|_F\|{\cal Z}\|_F(1-2h)$ gives a $2\epsilon_5\|\BS_{i_1=j}\|_F\|{\cal Z}\|_F$-approximate of ${\bf g}(\U_{i_1\bullet})$. Then, if we take $\epsilon=2\epsilon_5\|\BS_{i_1=j}\|_F\|{\cal Z}\|_F$, we can obtain the value of ${\bf g}_j(\U_{i_1\bullet})$ in time $O(m\log d\|\BS_{i_1=j}\|_F\|{\cal Z}\|_F/\epsilon)$ to accuracy $\epsilon$. Since the gradient ${\bf g}$ has $d$ entries, and we have to repeat the above procedure for all the singular matrices, then the total complexity of matrix optimization is $O(Km^{2}d{\rm polylog}d)$. Compared to the corresponding classical algorithm which takes $O(Kmd^m)$ classical calculations, our quantum algorithm is exponentially faster. To calculate the core tensor $\BS$, we can directly use the classical computation, the complexity is $O(Kd^m)$.

\section{Summary and Discussion}
We have introduced two quantum algorithms for higher order singular value decomposition. For the first method, the input has to be low-rank. For the second one, the input has no constraints on its structure. The output is a core tensor including tensor singular values and singular matrices stored in the quantum register. Some subroutines of the first method has been realized by quantum physicists, such as sparse Hamiltonian simulation and qPCA. For the second method, the QSVE algorithm is derived strictly by using quantum operators.

For an $m$th-order $n$-dimensional tensor, the complexity of the classical HOSVD is $O(m\,n^{m+1})$, while the complexities of our two quantum HOSVD algorithms for obtaining the superposition state of eigenvalues and eigenstates are $O(m^2 {\rm polylog}n)$ and $O(m^3 {\rm polylog}n)$ respectively. Moreover, the complexities of obtaining singular matrices and core tensor are $O(m^3n^{2}{\rm polylog} n)$ and $O(m^4n^{2}{\rm polylog} n)$ respectively. Usually, the order $m$ is much smaller than the dimension $n$. In this sense, our quantum HOSVD algorithms provide an exponential speedup over the classical counterpart.

Furthermore, we have provided an application of HOSVD on recommendation systems by quantum computers. Based on the given preference tensor, we may find out the missing values and give recommendations to users by this HOSVD model. One day the size of datasets is not enough for classical computers to handle, quantum computers will be able to solve these problems efficiently.

\section{Data Availability}
Data sharing is not applicable to this article as no datasets were generated or analysed during the current study.


\providecommand{\noopsort}[1]{}\providecommand{\singleletter}[1]{#1}%
\begin{thebibliography}{42}%
\makeatletter
\providecommand \@ifxundefined [1]{%
 \@ifx{#1\undefined}
}%
\providecommand \@ifnum [1]{%
 \ifnum #1\expandafter \@firstoftwo
 \else \expandafter \@secondoftwo
 \fi
}%
\providecommand \@ifx [1]{%
 \ifx #1\expandafter \@firstoftwo
 \else \expandafter \@secondoftwo
 \fi
}%
\providecommand \natexlab [1]{#1}%
\providecommand \enquote  [1]{``#1''}%
\providecommand \bibnamefont  [1]{#1}%
\providecommand \bibfnamefont [1]{#1}%
\providecommand \citenamefont [1]{#1}%
\providecommand \href@noop [0]{\@secondoftwo}%
\providecommand \href [0]{\begingroup \@sanitize@url \@href}%
\providecommand \@href[1]{\@@startlink{#1}\@@href}%
\providecommand \@@href[1]{\endgroup#1\@@endlink}%
\providecommand \@sanitize@url [0]{\catcode `\\12\catcode `\$12\catcode
  `\&12\catcode `\#12\catcode `\^12\catcode `\_12\catcode `\%12\relax}%
\providecommand \@@startlink[1]{}%
\providecommand \@@endlink[0]{}%
\providecommand \url  [0]{\begingroup\@sanitize@url \@url }%
\providecommand \@url [1]{\endgroup\@href {#1}{\urlprefix }}%
\providecommand \urlprefix  [0]{URL }%
\providecommand \Eprint [0]{\href }%
\providecommand \doibase [0]{https://doi.org/}%
\providecommand \selectlanguage [0]{\@gobble}%
\providecommand \bibinfo  [0]{\@secondoftwo}%
\providecommand \bibfield  [0]{\@secondoftwo}%
\providecommand \translation [1]{[#1]}%
\providecommand \BibitemOpen [0]{}%
\providecommand \bibitemStop [0]{}%
\providecommand \bibitemNoStop [0]{.\EOS\space}%
\providecommand \EOS [0]{\spacefactor3000\relax}%
\providecommand \BibitemShut  [1]{\csname bibitem#1\endcsname}%
\let\auto@bib@innerbib\@empty
\bibitem [{\citenamefont {Kilmer}\ and\ \citenamefont
  {Martin}(2011)}]{t-svd_three}%
  \BibitemOpen
  \bibfield  {author} {\bibinfo {author} {\bibfnamefont {M.~E.}\ \bibnamefont
  {Kilmer}}\ and\ \bibinfo {author} {\bibfnamefont {C.~D.}\ \bibnamefont
  {Martin}},\ }\bibfield  {title} {\bibinfo {title} {Factorization strategies
  for third-order tensors},\ }\href@noop {} {\bibfield  {journal} {\bibinfo
  {journal} {Linear Algebra and its Applications}\ }\textbf {\bibinfo {volume}
  {435}},\ \bibinfo {pages} {641} (\bibinfo {year} {2011})}\BibitemShut
  {NoStop}%
\bibitem [{\citenamefont {Zhang}\ and\ \citenamefont
  {Aeron}(2016)}]{exact_tensor_completion}%
  \BibitemOpen
  \bibfield  {author} {\bibinfo {author} {\bibfnamefont {Z.}~\bibnamefont
  {Zhang}}\ and\ \bibinfo {author} {\bibfnamefont {S.}~\bibnamefont {Aeron}},\
  }\bibfield  {title} {\bibinfo {title} {Exact tensor completion using t-svd},\
  }\href@noop {} {\bibfield  {journal} {\bibinfo  {journal} {IEEE Transactions
  on Signal Processing}\ }\textbf {\bibinfo {volume} {65}},\ \bibinfo {pages}
  {1511} (\bibinfo {year} {2016})}\BibitemShut {NoStop}%
\bibitem [{\citenamefont {Zhang}\ \emph {et~al.}(2014)\citenamefont {Zhang},
  \citenamefont {Ely}, \citenamefont {Aeron}, \citenamefont {Hao},\ and\
  \citenamefont {Kilmer}}]{Novel}%
  \BibitemOpen
  \bibfield  {author} {\bibinfo {author} {\bibfnamefont {Z.}~\bibnamefont
  {Zhang}}, \bibinfo {author} {\bibfnamefont {G.}~\bibnamefont {Ely}}, \bibinfo
  {author} {\bibfnamefont {S.}~\bibnamefont {Aeron}}, \bibinfo {author}
  {\bibfnamefont {N.}~\bibnamefont {Hao}},\ and\ \bibinfo {author}
  {\bibfnamefont {M.}~\bibnamefont {Kilmer}},\ }\bibfield  {title} {\bibinfo
  {title} {Novel methods for multilinear data completion and de-noising based
  on tensor-svd},\ }in\ \href@noop {} {\emph {\bibinfo {booktitle} {Proceedings
  of the IEEE conference on computer vision and pattern recognition}}}\
  (\bibinfo {year} {2014})\ pp.\ \bibinfo {pages} {3842--3849}\BibitemShut
  {NoStop}%
\bibitem [{\citenamefont {Ely}\ \emph {et~al.}(2015)\citenamefont {Ely},
  \citenamefont {Aeron}, \citenamefont {Hao},\ and\ \citenamefont
  {Kilmer}}]{5D}%
  \BibitemOpen
  \bibfield  {author} {\bibinfo {author} {\bibfnamefont {G.}~\bibnamefont
  {Ely}}, \bibinfo {author} {\bibfnamefont {S.}~\bibnamefont {Aeron}}, \bibinfo
  {author} {\bibfnamefont {N.}~\bibnamefont {Hao}},\ and\ \bibinfo {author}
  {\bibfnamefont {M.~E.}\ \bibnamefont {Kilmer}},\ }\bibfield  {title}
  {\bibinfo {title} {5d seismic data completion and denoising using a novel
  class of tensor decompositions},\ }\href@noop {} {\bibfield  {journal}
  {\bibinfo  {journal} {Geophysics}\ }\textbf {\bibinfo {volume} {80}},\
  \bibinfo {pages} {V83} (\bibinfo {year} {2015})}\BibitemShut {NoStop}%
\bibitem [{\citenamefont {Zhou}\ \emph {et~al.}(2017)\citenamefont {Zhou},
  \citenamefont {Lu}, \citenamefont {Lin},\ and\ \citenamefont
  {Zhang}}]{Tensor_factorization_for_low-rank}%
  \BibitemOpen
  \bibfield  {author} {\bibinfo {author} {\bibfnamefont {P.}~\bibnamefont
  {Zhou}}, \bibinfo {author} {\bibfnamefont {C.}~\bibnamefont {Lu}}, \bibinfo
  {author} {\bibfnamefont {Z.}~\bibnamefont {Lin}},\ and\ \bibinfo {author}
  {\bibfnamefont {C.}~\bibnamefont {Zhang}},\ }\bibfield  {title} {\bibinfo
  {title} {Tensor factorization for low-rank tensor completion},\ }\href@noop
  {} {\bibfield  {journal} {\bibinfo  {journal} {IEEE Transactions on Image
  Processing}\ }\textbf {\bibinfo {volume} {27}},\ \bibinfo {pages} {1152}
  (\bibinfo {year} {2017})}\BibitemShut {NoStop}%
\bibitem [{\citenamefont {Kolda}\ and\ \citenamefont {Bader}(2009)}]{KB09}%
  \BibitemOpen
  \bibfield  {author} {\bibinfo {author} {\bibfnamefont {T.~G.}\ \bibnamefont
  {Kolda}}\ and\ \bibinfo {author} {\bibfnamefont {B.~W.}\ \bibnamefont
  {Bader}},\ }\bibfield  {title} {\bibinfo {title} {Tensor decompositions and
  applications},\ }\href@noop {} {\bibfield  {journal} {\bibinfo  {journal}
  {SIAM review}\ }\textbf {\bibinfo {volume} {51}},\ \bibinfo {pages} {455}
  (\bibinfo {year} {2009})}\BibitemShut {NoStop}%
\bibitem [{\citenamefont {Qi}\ and\ \citenamefont {Luo}(2017)}]{Qi}%
  \BibitemOpen
  \bibfield  {author} {\bibinfo {author} {\bibfnamefont {L.}~\bibnamefont
  {Qi}}\ and\ \bibinfo {author} {\bibfnamefont {Z.}~\bibnamefont {Luo}},\
  }\href@noop {} {\emph {\bibinfo {title} {Tensor analysis: spectral theory and
  special tensors}}},\ Vol.\ \bibinfo {volume} {151}\ (\bibinfo  {publisher}
  {Siam},\ \bibinfo {year} {2017})\BibitemShut {NoStop}%
\bibitem [{\citenamefont {Qi}\ \emph {et~al.}(2018{\natexlab{a}})\citenamefont
  {Qi}, \citenamefont {Chen},\ and\ \citenamefont {Chen}}]{Qi2}%
  \BibitemOpen
  \bibfield  {author} {\bibinfo {author} {\bibfnamefont {L.}~\bibnamefont
  {Qi}}, \bibinfo {author} {\bibfnamefont {H.}~\bibnamefont {Chen}},\ and\
  \bibinfo {author} {\bibfnamefont {Y.}~\bibnamefont {Chen}},\ }\href@noop {}
  {\emph {\bibinfo {title} {Tensor eigenvalues and their applications}}},\
  Vol.~\bibinfo {volume} {39}\ (\bibinfo  {publisher} {Springer},\ \bibinfo
  {year} {2018})\BibitemShut {NoStop}%
\bibitem [{\citenamefont {Rendle}\ and\ \citenamefont
  {Schmidt-Thieme}(2010)}]{rendle2010pairwise}%
  \BibitemOpen
  \bibfield  {author} {\bibinfo {author} {\bibfnamefont {S.}~\bibnamefont
  {Rendle}}\ and\ \bibinfo {author} {\bibfnamefont {L.}~\bibnamefont
  {Schmidt-Thieme}},\ }\bibfield  {title} {\bibinfo {title} {Pairwise
  interaction tensor factorization for personalized tag recommendation},\ }in\
  \href@noop {} {\emph {\bibinfo {booktitle} {Proceedings of the third ACM
  international conference on Web search and data mining}}}\ (\bibinfo {year}
  {2010})\ pp.\ \bibinfo {pages} {81--90}\BibitemShut {NoStop}%
\bibitem [{\citenamefont {Rendle}\ \emph {et~al.}(2009)\citenamefont {Rendle},
  \citenamefont {Balby~Marinho}, \citenamefont {Nanopoulos},\ and\
  \citenamefont {Schmidt-Thieme}}]{rendle2009learning}%
  \BibitemOpen
  \bibfield  {author} {\bibinfo {author} {\bibfnamefont {S.}~\bibnamefont
  {Rendle}}, \bibinfo {author} {\bibfnamefont {L.}~\bibnamefont
  {Balby~Marinho}}, \bibinfo {author} {\bibfnamefont {A.}~\bibnamefont
  {Nanopoulos}},\ and\ \bibinfo {author} {\bibfnamefont {L.}~\bibnamefont
  {Schmidt-Thieme}},\ }\bibfield  {title} {\bibinfo {title} {Learning optimal
  ranking with tensor factorization for tag recommendation},\ }in\ \href@noop
  {} {\emph {\bibinfo {booktitle} {Proceedings of the 15th ACM SIGKDD
  international conference on Knowledge discovery and data mining}}}\ (\bibinfo
  {year} {2009})\ pp.\ \bibinfo {pages} {727--736}\BibitemShut {NoStop}%
\bibitem [{\citenamefont {Hu}\ \emph {et~al.}(2016)\citenamefont {Hu},
  \citenamefont {Qi},\ and\ \citenamefont {Zhang}}]{HQZ16}%
  \BibitemOpen
  \bibfield  {author} {\bibinfo {author} {\bibfnamefont {S.}~\bibnamefont
  {Hu}}, \bibinfo {author} {\bibfnamefont {L.}~\bibnamefont {Qi}},\ and\
  \bibinfo {author} {\bibfnamefont {G.}~\bibnamefont {Zhang}},\ }\bibfield
  {title} {\bibinfo {title} {Computing the geometric measure of entanglement of
  multipartite pure states by means of non-negative tensors},\ }\href@noop {}
  {\bibfield  {journal} {\bibinfo  {journal} {Physical Review A}\ }\textbf
  {\bibinfo {volume} {93}},\ \bibinfo {pages} {012304} (\bibinfo {year}
  {2016})}\BibitemShut {NoStop}%
\bibitem [{\citenamefont {Zhang}(2017)}]{Z17}%
  \BibitemOpen
  \bibfield  {author} {\bibinfo {author} {\bibfnamefont {G.}~\bibnamefont
  {Zhang}},\ }\bibfield  {title} {\bibinfo {title} {Dynamical analysis of
  quantum linear systems driven by multi-channel multi-photon states},\
  }\href@noop {} {\bibfield  {journal} {\bibinfo  {journal} {Automatica}\
  }\textbf {\bibinfo {volume} {83}},\ \bibinfo {pages} {186} (\bibinfo {year}
  {2017})}\BibitemShut {NoStop}%
\bibitem [{\citenamefont {Qi}\ \emph {et~al.}(2017)\citenamefont {Qi},
  \citenamefont {Zhang}, \citenamefont {Braun}, \citenamefont
  {Bohnet-Waldraff},\ and\ \citenamefont {Giraud}}]{QZB17}%
  \BibitemOpen
  \bibfield  {author} {\bibinfo {author} {\bibfnamefont {L.}~\bibnamefont
  {Qi}}, \bibinfo {author} {\bibfnamefont {G.}~\bibnamefont {Zhang}}, \bibinfo
  {author} {\bibfnamefont {D.}~\bibnamefont {Braun}}, \bibinfo {author}
  {\bibfnamefont {F.}~\bibnamefont {Bohnet-Waldraff}},\ and\ \bibinfo {author}
  {\bibfnamefont {O.}~\bibnamefont {Giraud}},\ }\bibfield  {title} {\bibinfo
  {title} {Regularly decomposable tensors and classical spin states},\
  }\href@noop {} {\bibfield  {journal} {\bibinfo  {journal} {Communications in
  mathematical sciences}\ } (\bibinfo {year} {2017})}\BibitemShut {NoStop}%
\bibitem [{\citenamefont {Qi}\ \emph {et~al.}(2018{\natexlab{b}})\citenamefont
  {Qi}, \citenamefont {Zhang},\ and\ \citenamefont {Ni}}]{QZN18}%
  \BibitemOpen
  \bibfield  {author} {\bibinfo {author} {\bibfnamefont {L.}~\bibnamefont
  {Qi}}, \bibinfo {author} {\bibfnamefont {G.}~\bibnamefont {Zhang}},\ and\
  \bibinfo {author} {\bibfnamefont {G.}~\bibnamefont {Ni}},\ }\bibfield
  {title} {\bibinfo {title} {How entangled can a multi-party system possibly
  be?},\ }\href@noop {} {\bibfield  {journal} {\bibinfo  {journal} {Physics
  Letters A}\ }\textbf {\bibinfo {volume} {382}},\ \bibinfo {pages} {1465}
  (\bibinfo {year} {2018}{\natexlab{b}})}\BibitemShut {NoStop}%
\bibitem [{\citenamefont {Rebentrost}\ \emph {et~al.}(2019)\citenamefont
  {Rebentrost}, \citenamefont {Schuld}, \citenamefont {Wossnig}, \citenamefont
  {Petruccione},\ and\ \citenamefont {Lloyd}}]{PMW19}%
  \BibitemOpen
  \bibfield  {author} {\bibinfo {author} {\bibfnamefont {P.}~\bibnamefont
  {Rebentrost}}, \bibinfo {author} {\bibfnamefont {M.}~\bibnamefont {Schuld}},
  \bibinfo {author} {\bibfnamefont {L.}~\bibnamefont {Wossnig}}, \bibinfo
  {author} {\bibfnamefont {F.}~\bibnamefont {Petruccione}},\ and\ \bibinfo
  {author} {\bibfnamefont {S.}~\bibnamefont {Lloyd}},\ }\bibfield  {title}
  {\bibinfo {title} {Quantum gradient descent and newton's method for
  constrained polynomial optimization},\ }\href@noop {} {\bibfield  {journal}
  {\bibinfo  {journal} {New Journal of Physics}\ }\textbf {\bibinfo {volume}
  {21}},\ \bibinfo {pages} {073023} (\bibinfo {year} {2019})}\BibitemShut
  {NoStop}%
\bibitem [{\citenamefont {Zhang}\ \emph {et~al.}(2019)\citenamefont {Zhang},
  \citenamefont {Ni},\ and\ \citenamefont {Zhang}}]{ZNZ19}%
  \BibitemOpen
  \bibfield  {author} {\bibinfo {author} {\bibfnamefont {M.}~\bibnamefont
  {Zhang}}, \bibinfo {author} {\bibfnamefont {G.}~\bibnamefont {Ni}},\ and\
  \bibinfo {author} {\bibfnamefont {G.}~\bibnamefont {Zhang}},\ }\bibfield
  {title} {\bibinfo {title} {Iterative methods for computing u-eigenvalues of
  non-symmetric complex tensors with application in quantum entanglement},\
  }\href@noop {} {\bibfield  {journal} {\bibinfo  {journal} {Computational
  Optimization and Applications}\ ,\ \bibinfo {pages} {1}} (\bibinfo {year}
  {2019})}\BibitemShut {NoStop}%
\bibitem [{\citenamefont {Wu}\ \emph {et~al.}(2019)\citenamefont {Wu},
  \citenamefont {He}, \citenamefont {Zhang}, \citenamefont {Chen},
  \citenamefont {Sun}, \citenamefont {Liu}, \citenamefont {Zhang},\ and\
  \citenamefont {Poor}}]{WHZ19}%
  \BibitemOpen
  \bibfield  {author} {\bibinfo {author} {\bibfnamefont {M.}~\bibnamefont
  {Wu}}, \bibinfo {author} {\bibfnamefont {S.}~\bibnamefont {He}}, \bibinfo
  {author} {\bibfnamefont {Y.}~\bibnamefont {Zhang}}, \bibinfo {author}
  {\bibfnamefont {J.}~\bibnamefont {Chen}}, \bibinfo {author} {\bibfnamefont
  {Y.}~\bibnamefont {Sun}}, \bibinfo {author} {\bibfnamefont {Y.-Y.}\
  \bibnamefont {Liu}}, \bibinfo {author} {\bibfnamefont {J.}~\bibnamefont
  {Zhang}},\ and\ \bibinfo {author} {\bibfnamefont {H.~V.}\ \bibnamefont
  {Poor}},\ }\bibfield  {title} {\bibinfo {title} {A tensor-based framework for
  studying eigenvector multicentrality in multilayer networks},\ }\href@noop {}
  {\bibfield  {journal} {\bibinfo  {journal} {Proceedings of the National
  Academy of Sciences}\ }\textbf {\bibinfo {volume} {116}},\ \bibinfo {pages}
  {15407} (\bibinfo {year} {2019})}\BibitemShut {NoStop}%
\bibitem [{\citenamefont {Huggins}\ \emph {et~al.}(2019)\citenamefont
  {Huggins}, \citenamefont {Patil}, \citenamefont {Mitchell}, \citenamefont
  {Whaley},\ and\ \citenamefont {Stoudenmire}}]{HPM19}%
  \BibitemOpen
  \bibfield  {author} {\bibinfo {author} {\bibfnamefont {W.}~\bibnamefont
  {Huggins}}, \bibinfo {author} {\bibfnamefont {P.}~\bibnamefont {Patil}},
  \bibinfo {author} {\bibfnamefont {B.}~\bibnamefont {Mitchell}}, \bibinfo
  {author} {\bibfnamefont {K.~B.}\ \bibnamefont {Whaley}},\ and\ \bibinfo
  {author} {\bibfnamefont {E.~M.}\ \bibnamefont {Stoudenmire}},\ }\bibfield
  {title} {\bibinfo {title} {Towards quantum machine learning with tensor
  networks},\ }\href@noop {} {\bibfield  {journal} {\bibinfo  {journal}
  {Quantum Science and technology}\ }\textbf {\bibinfo {volume} {4}},\ \bibinfo
  {pages} {024001} (\bibinfo {year} {2019})}\BibitemShut {NoStop}%
\bibitem [{\citenamefont {Ma}\ \emph {et~al.}(2020)\citenamefont {Ma},
  \citenamefont {Wang},\ and\ \citenamefont {Tresp}}]{MWT20}%
  \BibitemOpen
  \bibfield  {author} {\bibinfo {author} {\bibfnamefont {Y.}~\bibnamefont
  {Ma}}, \bibinfo {author} {\bibfnamefont {Y.}~\bibnamefont {Wang}},\ and\
  \bibinfo {author} {\bibfnamefont {V.}~\bibnamefont {Tresp}},\ }\bibfield
  {title} {\bibinfo {title} {Quantum machine learning algorithm for knowledge
  graphs},\ }\href@noop {} {\bibfield  {journal} {\bibinfo  {journal} {arXiv
  preprint arXiv:2001.01077}\ } (\bibinfo {year} {2020})}\BibitemShut {NoStop}%
\bibitem [{\citenamefont {Comon}(2002)}]{CP}%
  \BibitemOpen
  \bibfield  {author} {\bibinfo {author} {\bibfnamefont {P.}~\bibnamefont
  {Comon}},\ }\bibfield  {title} {\bibinfo {title} {Tensor decompositions},\
  }\href@noop {} {\bibfield  {journal} {\bibinfo  {journal} {Mathematics in
  Signal Processing V}\ ,\ \bibinfo {pages} {1}} (\bibinfo {year}
  {2002})}\BibitemShut {NoStop}%
\bibitem [{\citenamefont {Tucker}(1966)}]{Tucker}%
  \BibitemOpen
  \bibfield  {author} {\bibinfo {author} {\bibfnamefont {L.~R.}\ \bibnamefont
  {Tucker}},\ }\bibfield  {title} {\bibinfo {title} {Some mathematical notes on
  three-mode factor analysis},\ }\href@noop {} {\bibfield  {journal} {\bibinfo
  {journal} {Psychometrika}\ }\textbf {\bibinfo {volume} {31}},\ \bibinfo
  {pages} {279} (\bibinfo {year} {1966})}\BibitemShut {NoStop}%
\bibitem [{\citenamefont {De~Lathauwer}\ \emph {et~al.}(2000)\citenamefont
  {De~Lathauwer}, \citenamefont {De~Moor},\ and\ \citenamefont
  {Vandewalle}}]{HOSVD}%
  \BibitemOpen
  \bibfield  {author} {\bibinfo {author} {\bibfnamefont {L.}~\bibnamefont
  {De~Lathauwer}}, \bibinfo {author} {\bibfnamefont {B.}~\bibnamefont
  {De~Moor}},\ and\ \bibinfo {author} {\bibfnamefont {J.}~\bibnamefont
  {Vandewalle}},\ }\bibfield  {title} {\bibinfo {title} {A multilinear singular
  value decomposition},\ }\href@noop {} {\bibfield  {journal} {\bibinfo
  {journal} {SIAM journal on Matrix Analysis and Applications}\ }\textbf
  {\bibinfo {volume} {21}},\ \bibinfo {pages} {1253} (\bibinfo {year}
  {2000})}\BibitemShut {NoStop}%
\bibitem [{\citenamefont {{Gu}}\ \emph {et~al.}(2019)\citenamefont {{Gu}},
  \citenamefont {{Wang}},\ and\ \citenamefont {{Zhang}}}]{GWZ19}%
  \BibitemOpen
  \bibfield  {author} {\bibinfo {author} {\bibfnamefont {L.}~\bibnamefont
  {{Gu}}}, \bibinfo {author} {\bibfnamefont {X.}~\bibnamefont {{Wang}}},\ and\
  \bibinfo {author} {\bibfnamefont {G.}~\bibnamefont {{Zhang}}},\ }\bibfield
  {title} {\bibinfo {title} {Quantum higher order singular value
  decomposition},\ }in\ \href@noop {} {\emph {\bibinfo {booktitle} {2019 IEEE
  International Conference on Systems, Man and Cybernetics (SMC)}}}\ (\bibinfo
  {year} {2019})\ pp.\ \bibinfo {pages} {1166--1171}\BibitemShut {NoStop}%
\bibitem [{\citenamefont {Oseledets}(2011)}]{oseledets2011tensor}%
  \BibitemOpen
  \bibfield  {author} {\bibinfo {author} {\bibfnamefont {I.~V.}\ \bibnamefont
  {Oseledets}},\ }\bibfield  {title} {\bibinfo {title} {Tensor-train
  decomposition},\ }\href@noop {} {\bibfield  {journal} {\bibinfo  {journal}
  {SIAM Journal on Scientific Computing}\ }\textbf {\bibinfo {volume} {33}},\
  \bibinfo {pages} {2295} (\bibinfo {year} {2011})}\BibitemShut {NoStop}%
\bibitem [{\citenamefont {Or{\'u}s}(2014)}]{orus14}%
  \BibitemOpen
  \bibfield  {author} {\bibinfo {author} {\bibfnamefont {R.}~\bibnamefont
  {Or{\'u}s}},\ }\bibfield  {title} {\bibinfo {title} {A practical introduction
  to tensor networks: Matrix product states and projected entangled pair
  states},\ }\href@noop {} {\bibfield  {journal} {\bibinfo  {journal} {Annals
  of Physics}\ }\textbf {\bibinfo {volume} {349}},\ \bibinfo {pages} {117}
  (\bibinfo {year} {2014})}\BibitemShut {NoStop}%
\bibitem [{\citenamefont {Martin}\ \emph {et~al.}(2013)\citenamefont {Martin},
  \citenamefont {Shafer},\ and\ \citenamefont {LaRue}}]{order-p}%
  \BibitemOpen
  \bibfield  {author} {\bibinfo {author} {\bibfnamefont {C.~D.}\ \bibnamefont
  {Martin}}, \bibinfo {author} {\bibfnamefont {R.}~\bibnamefont {Shafer}},\
  and\ \bibinfo {author} {\bibfnamefont {B.}~\bibnamefont {LaRue}},\ }\bibfield
   {title} {\bibinfo {title} {An order-$p$ tensor factorization with
  applications in imaging},\ }\href@noop {} {\bibfield  {journal} {\bibinfo
  {journal} {SIAM Journal on Scientific Computing}\ }\textbf {\bibinfo {volume}
  {35}},\ \bibinfo {pages} {A474} (\bibinfo {year} {2013})}\BibitemShut
  {NoStop}%
\bibitem [{\citenamefont {Nielsen}\ and\ \citenamefont
  {Chuang}(2010)}]{Nielsen}%
  \BibitemOpen
  \bibfield  {author} {\bibinfo {author} {\bibfnamefont {M.}~\bibnamefont
  {Nielsen}}\ and\ \bibinfo {author} {\bibfnamefont {I.}~\bibnamefont
  {Chuang}},\ }\href@noop {} {\emph {\bibinfo {title} {Quantum Computation and
  Information}}}\ (\bibinfo  {publisher} {Cambridge University Press},\
  \bibinfo {address} {London},\ \bibinfo {year} {2010})\BibitemShut {NoStop}%
\bibitem [{\citenamefont {Kerenidis}\ and\ \citenamefont
  {Prakash}(2017{\natexlab{a}})}]{QRS}%
  \BibitemOpen
  \bibfield  {author} {\bibinfo {author} {\bibfnamefont {I.}~\bibnamefont
  {Kerenidis}}\ and\ \bibinfo {author} {\bibfnamefont {A.}~\bibnamefont
  {Prakash}},\ }\bibfield  {title} {\bibinfo {title} {{Quantum Recommendation
  Systems}},\ }in\ \href@noop {} {\emph {\bibinfo {booktitle} {8th Innovations
  in Theoretical Computer Science Conference (ITCS 2017)}}},\ \bibinfo {series}
  {Leibniz International Proceedings in Informatics (LIPIcs)}, Vol.~\bibinfo
  {volume} {67},\ \bibinfo {editor} {edited by\ \bibinfo {editor}
  {\bibfnamefont {C.~H.}\ \bibnamefont {Papadimitriou}}}\ (\bibinfo
  {publisher} {Schloss Dagstuhl--Leibniz-Zentrum fuer Informatik},\ \bibinfo
  {address} {Dagstuhl, Germany},\ \bibinfo {year} {2017})\ pp.\ \bibinfo
  {pages} {49:1--49:21}\BibitemShut {NoStop}%
\bibitem [{\citenamefont {Rebentrost}\ \emph {et~al.}(2018)\citenamefont
  {Rebentrost}, \citenamefont {Steffens}, \citenamefont {Marvian},\ and\
  \citenamefont {Lloyd}}]{QSVD}%
  \BibitemOpen
  \bibfield  {author} {\bibinfo {author} {\bibfnamefont {P.}~\bibnamefont
  {Rebentrost}}, \bibinfo {author} {\bibfnamefont {A.}~\bibnamefont
  {Steffens}}, \bibinfo {author} {\bibfnamefont {I.}~\bibnamefont {Marvian}},\
  and\ \bibinfo {author} {\bibfnamefont {S.}~\bibnamefont {Lloyd}},\ }\bibfield
   {title} {\bibinfo {title} {Quantum singular-value decomposition of nonsparse
  low-rank matrices},\ }\href@noop {} {\bibfield  {journal} {\bibinfo
  {journal} {Physical review A}\ }\textbf {\bibinfo {volume} {97}},\ \bibinfo
  {pages} {012327} (\bibinfo {year} {2018})}\BibitemShut {NoStop}%
\bibitem [{\citenamefont {Hao}\ \emph {et~al.}(2013{\natexlab{a}})\citenamefont
  {Hao}, \citenamefont {Kilmer}, \citenamefont {Braman},\ and\ \citenamefont
  {Hoover}}]{Hao2013Facial}%
  \BibitemOpen
  \bibfield  {author} {\bibinfo {author} {\bibfnamefont {N.}~\bibnamefont
  {Hao}}, \bibinfo {author} {\bibfnamefont {M.}~\bibnamefont {Kilmer}},
  \bibinfo {author} {\bibfnamefont {K.}~\bibnamefont {Braman}},\ and\ \bibinfo
  {author} {\bibfnamefont {R.}~\bibnamefont {Hoover}},\ }\bibfield  {title}
  {\bibinfo {title} {Facial recognition using tensor-tensor decompositions},\
  }\href {https://doi.org/10.1137/110842570} {\bibfield  {journal} {\bibinfo
  {journal} {SIAM Journal on Imaging Sciences [electronic only]}\ }\textbf
  {\bibinfo {volume} {6}} (\bibinfo {year} {2013}{\natexlab{a}})}\BibitemShut
  {NoStop}%
\bibitem [{\citenamefont {Teixeira}\ and\ \citenamefont
  {Rodriguez}(1995)}]{convolution}%
  \BibitemOpen
  \bibfield  {author} {\bibinfo {author} {\bibfnamefont {M.}~\bibnamefont
  {Teixeira}}\ and\ \bibinfo {author} {\bibfnamefont {D.}~\bibnamefont
  {Rodriguez}},\ }\bibfield  {title} {\bibinfo {title} {A class of fast cyclic
  convolution algorithms based on block pseudocirculants},\ }\href@noop {}
  {\bibfield  {journal} {\bibinfo  {journal} {IEEE Signal Processing Letters}\
  }\textbf {\bibinfo {volume} {2}},\ \bibinfo {pages} {92} (\bibinfo {year}
  {1995})}\BibitemShut {NoStop}%
\bibitem [{\citenamefont {Hao}\ \emph {et~al.}(2013{\natexlab{b}})\citenamefont
  {Hao}, \citenamefont {Kilmer}, \citenamefont {Braman},\ and\ \citenamefont
  {Hoover}}]{t-qr}%
  \BibitemOpen
  \bibfield  {author} {\bibinfo {author} {\bibfnamefont {N.}~\bibnamefont
  {Hao}}, \bibinfo {author} {\bibfnamefont {M.~E.}\ \bibnamefont {Kilmer}},
  \bibinfo {author} {\bibfnamefont {K.}~\bibnamefont {Braman}},\ and\ \bibinfo
  {author} {\bibfnamefont {R.~C.}\ \bibnamefont {Hoover}},\ }\bibfield  {title}
  {\bibinfo {title} {Facial recognition using tensor-tensor decompositions},\
  }\href@noop {} {\bibfield  {journal} {\bibinfo  {journal} {SIAM Journal on
  Imaging Sciences}\ }\textbf {\bibinfo {volume} {6}},\ \bibinfo {pages} {437}
  (\bibinfo {year} {2013}{\natexlab{b}})}\BibitemShut {NoStop}%
\bibitem [{\citenamefont {Zhang}\ \emph {et~al.}(2018)\citenamefont {Zhang},
  \citenamefont {Saibaba}, \citenamefont {Kilmer},\ and\ \citenamefont
  {Aeron}}]{randomized}%
  \BibitemOpen
  \bibfield  {author} {\bibinfo {author} {\bibfnamefont {J.}~\bibnamefont
  {Zhang}}, \bibinfo {author} {\bibfnamefont {A.~K.}\ \bibnamefont {Saibaba}},
  \bibinfo {author} {\bibfnamefont {M.~E.}\ \bibnamefont {Kilmer}},\ and\
  \bibinfo {author} {\bibfnamefont {S.}~\bibnamefont {Aeron}},\ }\bibfield
  {title} {\bibinfo {title} {A randomized tensor singular value decomposition
  based on the t-product},\ }\href@noop {} {\bibfield  {journal} {\bibinfo
  {journal} {Numerical Linear Algebra with Applications}\ }\textbf {\bibinfo
  {volume} {25}},\ \bibinfo {pages} {e2179} (\bibinfo {year}
  {2018})}\BibitemShut {NoStop}%
\bibitem [{\citenamefont {Lloyd}\ \emph {et~al.}(2014)\citenamefont {Lloyd},
  \citenamefont {Mohseni},\ and\ \citenamefont {Rebentrost}}]{QPCA}%
  \BibitemOpen
  \bibfield  {author} {\bibinfo {author} {\bibfnamefont {S.}~\bibnamefont
  {Lloyd}}, \bibinfo {author} {\bibfnamefont {M.}~\bibnamefont {Mohseni}},\
  and\ \bibinfo {author} {\bibfnamefont {P.}~\bibnamefont {Rebentrost}},\
  }\bibfield  {title} {\bibinfo {title} {Quantum principal component
  analysis},\ }\href@noop {} {\bibfield  {journal} {\bibinfo  {journal} {Nature
  Physics}\ }\textbf {\bibinfo {volume} {10}},\ \bibinfo {pages} {631}
  (\bibinfo {year} {2014})}\BibitemShut {NoStop}%
\bibitem [{\citenamefont {Achlioptas}\ and\ \citenamefont
  {McSherry}(2007)}]{Achlioptas}%
  \BibitemOpen
  \bibfield  {author} {\bibinfo {author} {\bibfnamefont {D.}~\bibnamefont
  {Achlioptas}}\ and\ \bibinfo {author} {\bibfnamefont {F.}~\bibnamefont
  {McSherry}},\ }\bibfield  {title} {\bibinfo {title} {Fast computation of
  low-rank matrix approximations},\ }\href@noop {} {\bibfield  {journal}
  {\bibinfo  {journal} {Journal of the ACM (JACM)}\ }\textbf {\bibinfo {volume}
  {54}},\ \bibinfo {pages} {9} (\bibinfo {year} {2007})}\BibitemShut {NoStop}%
\bibitem [{\citenamefont {Ross}(2006)}]{R06}%
  \BibitemOpen
  \bibfield  {author} {\bibinfo {author} {\bibfnamefont {S.~M.}\ \bibnamefont
  {Ross}},\ }\href@noop {} {\emph {\bibinfo {title} {Introduction to
  Probability Models, ISE}}}\ (\bibinfo  {publisher} {Academic press},\
  \bibinfo {year} {2006})\BibitemShut {NoStop}%
\bibitem [{\citenamefont {Adomavicius}\ and\ \citenamefont
  {Tuzhilin}(2011)}]{Context-aware}%
  \BibitemOpen
  \bibfield  {author} {\bibinfo {author} {\bibfnamefont {G.}~\bibnamefont
  {Adomavicius}}\ and\ \bibinfo {author} {\bibfnamefont {A.}~\bibnamefont
  {Tuzhilin}},\ }\bibfield  {title} {\bibinfo {title} {Context-aware
  recommender systems},\ }in\ \href@noop {} {\emph {\bibinfo {booktitle}
  {Recommender systems handbook}}}\ (\bibinfo  {publisher} {Springer},\
  \bibinfo {year} {2011})\ pp.\ \bibinfo {pages} {217--253}\BibitemShut
  {NoStop}%
\bibitem [{\citenamefont {Kerenidis}\ and\ \citenamefont
  {Prakash}(2017{\natexlab{b}})}]{kerenidis2017quantum}%
  \BibitemOpen
  \bibfield  {author} {\bibinfo {author} {\bibfnamefont {I.}~\bibnamefont
  {Kerenidis}}\ and\ \bibinfo {author} {\bibfnamefont {A.}~\bibnamefont
  {Prakash}},\ }\bibfield  {title} {\bibinfo {title} {Quantum gradient descent
  for linear systems and least squares},\ }\href@noop {} {\bibfield  {journal}
  {\bibinfo  {journal} {arXiv preprint arXiv:1704.04992}\ } (\bibinfo {year}
  {2017}{\natexlab{b}})}\BibitemShut {NoStop}%
\bibitem [{\citenamefont {Miao}\ \emph {et~al.}(2020)\citenamefont {Miao},
  \citenamefont {Qi},\ and\ \citenamefont {Wei}}]{Miao}%
  \BibitemOpen
  \bibfield  {author} {\bibinfo {author} {\bibfnamefont {Y.}~\bibnamefont
  {Miao}}, \bibinfo {author} {\bibfnamefont {L.}~\bibnamefont {Qi}},\ and\
  \bibinfo {author} {\bibfnamefont {Y.}~\bibnamefont {Wei}},\ }\bibfield
  {title} {\bibinfo {title} {Generalized tensor function via the tensor
  singular value decomposition based on the t-product},\ }\href@noop {}
  {\bibfield  {journal} {\bibinfo  {journal} {Linear Algebra and its
  Applications}\ }\textbf {\bibinfo {volume} {590}},\ \bibinfo {pages} {258}
  (\bibinfo {year} {2020})}\BibitemShut {NoStop}%
\bibitem [{\citenamefont {Prakash}(2014)}]{Prakash_Doc}%
  \BibitemOpen
  \bibfield  {author} {\bibinfo {author} {\bibfnamefont {A.}~\bibnamefont
  {Prakash}},\ }\emph {\bibinfo {title} {Quantum algorithms for linear algebra
  and machine learning.}},\ \href@noop {} {Ph.D. thesis},\ \bibinfo  {school}
  {UC Berkeley} (\bibinfo {year} {2014})\BibitemShut {NoStop}%
\bibitem [{\citenamefont {Giovannetti}\ \emph {et~al.}(2008)\citenamefont
  {Giovannetti}, \citenamefont {Lloyd},\ and\ \citenamefont {Maccone}}]{QRAM}%
  \BibitemOpen
  \bibfield  {author} {\bibinfo {author} {\bibfnamefont {V.}~\bibnamefont
  {Giovannetti}}, \bibinfo {author} {\bibfnamefont {S.}~\bibnamefont {Lloyd}},\
  and\ \bibinfo {author} {\bibfnamefont {L.}~\bibnamefont {Maccone}},\
  }\bibfield  {title} {\bibinfo {title} {Quantum random access memory},\
  }\href@noop {} {\bibfield  {journal} {\bibinfo  {journal} {Physical review
  letters}\ }\textbf {\bibinfo {volume} {100}},\ \bibinfo {pages} {160501}
  (\bibinfo {year} {2008})}\BibitemShut {NoStop}%
\bibitem [{\citenamefont {Shao}\ and\ \citenamefont
  {Xiang}(2018)}]{quantum_circulant}%
  \BibitemOpen
  \bibfield  {author} {\bibinfo {author} {\bibfnamefont {C.}~\bibnamefont
  {Shao}}\ and\ \bibinfo {author} {\bibfnamefont {H.}~\bibnamefont {Xiang}},\
  }\bibfield  {title} {\bibinfo {title} {Quantum circulant preconditioner for a
  linear system of equations},\ }\href@noop {} {\bibfield  {journal} {\bibinfo
  {journal} {Physical Review A}\ }\textbf {\bibinfo {volume} {98}},\ \bibinfo
  {pages} {062321} (\bibinfo {year} {2018})}\BibitemShut {NoStop}%
\end{thebibliography}%


\providecommand{\noopsort}[1]{}\providecommand{\singleletter}[1]{#1}%
%


\begin{thebibliography}{10}

\bibitem{AA}
Andris Ambainis.
\newblock Variable time amplitude amplification and quantum algorithms for
  linear algebra problems.
\newblock In {\em STACS'12 (29th Symposium on Theoretical Aspects of Computer
  Science)}, volume~14, pages 636--647. LIPIcs, 2012.

\bibitem{QCh}
Franziska Bell, Daniel~S Lambrecht, and Martin Head-Gordon.
\newblock Higher order singular value decomposition in quantum chemistry.
\newblock {\em Molecular Physics}, 108(19-20):2759--2773, 2010.

\bibitem{NF}
James Bennett and Stan Lanning.
\newblock The {N}etflix {P}rize.
\newblock In {\em Proceedings of KDD cup and workshop}, volume 2007, page~35.
  New York, NY, USA., 2007.

\bibitem{ESSH}
Dominic~W Berry, Graeme Ahokas, Richard Cleve, and Barry~C Sanders.
\newblock Efficient quantum algorithms for simulating sparse {H}amiltonians.
\newblock {\em Communications in Mathematical Physics}, 270(2):359--371, 2007.

\bibitem{QTN}
Jacob Biamonte and Ville Bergholm.
\newblock Tensor networks in a nutshell.
\newblock {\em arXiv preprint arXiv:1708.00006}, 2017.

\bibitem{SGD}
L{\'e}on Bottou.
\newblock Large-scale machine learning with stochastic gradient descent.
\newblock In {\em Proceedings of COMPSTAT'2010}, pages 177--186. Springer,
  2010.

\bibitem{QAE}
Gilles Brassard, Peter Hoyer, Michele Mosca, and Alain Tapp.
\newblock Quantum amplitude amplification and estimation.
\newblock {\em Contemporary Mathematics}, 305:53--74, 2002.

\bibitem{CWS+13}
X-D Cai, Christian Weedbrook, Z-E Su, M-C Chen, Mile Gu, M-J Zhu, Li~Li, Nai-Le
  Liu, Chao-Yang Lu, and Jian-Wei Pan.
\newblock Experimental quantum computing to solve systems of linear equations.
\newblock {\em Physical review letters}, 110(23):230501, 2013.

\bibitem{CP1}
J~Douglas Carroll and Jih-Jie Chang.
\newblock Analysis of individual differences in multidimensional scaling via an
  {N}-way generalization of ``{E}ckart-{Y}oung" decomposition.
\newblock {\em Psychometrika}, 35(3):283--319, 1970.

\bibitem{Lie2000}
Lieven De~Lathauwer, Bart De~Moor, and Joos Vandewalle.
\newblock A multilinear singular value decomposition.
\newblock {\em SIAM journal on Matrix Analysis and Applications},
  21(4):1253--1278, 2000.

\bibitem{MLQD}
Vedran Dunjko and Hans~J Briegel.
\newblock Machine learning \& artificial intelligence in the quantum domain: a
  review of recent progress.
\newblock {\em Reports on Progress in Physics}, 81(7):074001, 2018.

\bibitem{QRAM}
Vittorio Giovannetti, Seth Lloyd, and Lorenzo Maccone.
\newblock Quantum random access memory.
\newblock {\em Physical review letters}, 100(16):160501, 2008.

\bibitem{Grover}
Lov~K Grover.
\newblock A fast quantum mechanical algorithm for database search.
\newblock In {\em Proceedings of the Twenty-eighth Annual ACM Symposium on
  Theory of Computing}, pages 212--219. New York, NY, USA., 1996.

\bibitem{GWZ19}
Lejia Gu, Xiaoqiang Wang, and Guofeng Zhang.
\newblock Quantum higher order singular value decomposition.
\newblock In {\em 2019 IEEE International Conference on Systems, Man and
  Cybernetics (SMC)}, pages 1166--1171. IEEE, 2019.

\bibitem{HHL}
Aram~W Harrow, Avinatan Hassidim, and Seth Lloyd.
\newblock Quantum algorithm for linear systems of equations.
\newblock {\em Physical review letters}, 103(15):150502, 2009.

\bibitem{CP2}
Richard~A Harshman et~al.
\newblock Foundations of the {PARAFAC} procedure: Models and conditions for an
  ``explanatory" multimodal factor analysis.
\newblock 1970.

\bibitem{HQZ16}
Shenglong Hu, Liqun Qi, and Guofeng Zhang.
\newblock Computing the geometric measure of entanglement of multipartite pure
  states by means of non-negative tensors.
\newblock {\em Physical Review A}, 93(1):012304, 2016.

\bibitem{TQML}
William Huggins, Piyush Patil, Bradley Mitchell, K~Birgitta Whaley, and E~Miles
  Stoudenmire.
\newblock Towards quantum machine learning with tensor networks.
\newblock {\em Quantum Science and technology}, 4(2):024001, 2019.

\bibitem{Mulre}
Alexandros Karatzoglou, Xavier Amatriain, Linas Baltrunas, and Nuria Oliver.
\newblock Multiverse recommendation: n-dimensional tensor factorization for
  context-aware collaborative filtering.
\newblock In {\em Proceedings of the fourth ACM conference on Recommender
  systems}, pages 79--86. ACM, 2010.

\bibitem{QRS}
Iordanis Kerenidis and Anupam Prakash.
\newblock Quantum recommendation systems.
\newblock In Christos~H. Papadimitriou, editor, {\em 8th Innovations in
  Theoretical Computer Science Conference (ITCS 2017)}, volume~67 of {\em
  Leibniz International Proceedings in Informatics (LIPIcs)}, pages
  49:1--49:21, Dagstuhl, Germany, 2017. Schloss Dagstuhl--Leibniz-Zentrum fuer
  Informatik.

\bibitem{CNN}
Maksym Kholiavchenko.
\newblock Iterative low-rank approximation for {CNN} compression.
\newblock {\em arXiv preprint arXiv:1803.08995}, 2018.

\bibitem{Kitaev}
A~Yu Kitaev.
\newblock Quantum measurements and the {A}belian stabilizer problem.
\newblock {\em arXiv preprint quant-ph/9511026}, 1995.

\bibitem{KB09}
Tamara~G Kolda and Brett~W Bader.
\newblock Tensor decompositions and applications.
\newblock {\em SIAM review}, 51(3):455--500, 2009.

\bibitem{QPCA}
Seth Lloyd, Masoud Mohseni, and Patrick Rebentrost.
\newblock Quantum principal component analysis.
\newblock {\em Nature Physics}, 10(9):631, 2014.

\bibitem{Bible}
Michael~A Nielsen and Isaac Chuang.
\newblock Quantum computation and quantum information, 2002.

\bibitem{DNA}
Larsson Omberg, Gene~H Golub, and Orly Alter.
\newblock A tensor higher-order singular value decomposition for integrative
  analysis of {DNA} microarray data from different studies.
\newblock {\em Proceedings of the National Academy of Sciences},
  104(47):18371--18376, 2007.

\bibitem{TTD}
Ivan~V Oseledets.
\newblock Tensor-train decomposition.
\newblock {\em SIAM Journal on Scientific Computing}, 33(5):2295--2317, 2011.

\bibitem{PCY+14}
Jian Pan, Yudong Cao, Xiwei Yao, Zhaokai Li, Chenyong Ju, Hongwei Chen, Xinhua
  Peng, Sabre Kais, and Jiangfeng Du.
\newblock Experimental realization of quantum algorithm for solving linear
  systems of equations.
\newblock {\em Physical Review A}, 89(2):022313, 2014.

\bibitem{Pra}
Anupam Prakash.
\newblock {\em Quantum algorithms for linear algebra and machine learning}.
\newblock PhD thesis, UC Berkeley, 2014.

\bibitem{QCC}
Liqun Qi, Haibin Chen, and Yannan Chen.
\newblock {\em Tensor eigenvalues and their applications}, volume~39.
\newblock Springer, 2018.

\bibitem{Qi}
Liqun Qi and Ziyan Luo.
\newblock {\em Tensor analysis: spectral theory and special tensors}, volume
  151.
\newblock SIAM, 2017.

\bibitem{QZB17}
Liqun Qi, Guofeng Zhang, Daniel Braun, Fabian Bohnet-Waldraff, and Olivier
  Giraud.
\newblock Regularly decomposable tensors and classical spin states.
\newblock {\em Communications in Mathematical Sciences}, 15(6):1651--1665,
  2017.

\bibitem{QZN18}
Liqun Qi, Guofeng Zhang, and Guyan Ni.
\newblock How entangled can a multi-party system possibly be?
\newblock {\em Physics Letters A}, 382(22):1465--1471, 2018.

\bibitem{QSVM}
Patrick Rebentrost, Masoud Mohseni, and Seth Lloyd.
\newblock Quantum support vector machine for big data classification.
\newblock {\em Physical review letters}, 113(13):130503, 2014.

\bibitem{QSVD}
Patrick Rebentrost, Adrian Steffens, Iman Marvian, and Seth Lloyd.
\newblock Quantum singular-value decomposition of nonsparse low-rank matrices.
\newblock {\em Physical review A}, 97(1):012327, 2018.

\bibitem{NN}
Andr{\'a}s R{\"o}vid, L{\'a}szl{\'o} Szeidl, and P{\'e}ter V{\'a}rlaki.
\newblock On tensor-product model based representation of neural networks.
\newblock In {\em 2011 15th IEEE International Conference on Intelligent
  Engineering Systems}, pages 69--72. IEEE, 2011.

\bibitem{QMM}
Changpeng Shao.
\newblock Quantum algorithms to matrix multiplication.
\newblock {\em arXiv preprint arXiv:1803.01601}, 2018.

\bibitem{Shor}
Peter~W Shor.
\newblock Algorithms for quantum computation: discrete logarithms and
  factoring.
\newblock In {\em Proceedings 35th annual symposium on foundations of computer
  science}, pages 124--134. IEEE, 1994.

\bibitem{TC}
Qingquan Song, Hancheng Ge, James Caverlee, and Xia Hu.
\newblock Tensor completion algorithms in big data analytics.
\newblock {\em ACM Transactions on Knowledge Discovery from Data (TKDD)},
  13(1):1--48, 2019.

\bibitem{Tagre}
Panagiotis Symeonidis, Alexandros Nanopoulos, and Yannis Manolopoulos.
\newblock Tag recommendations based on tensor dimensionality reduction.
\newblock In {\em Proceedings of the 2008 ACM conference on Recommender
  systems}, pages 43--50. ACM, 2008.

\bibitem{nrhosvd}
L{\'a}szl{\'o} Szeidl, Peter Baranyi, Zolt{\'a}n Petres, and Peter Varlaki.
\newblock Numerical reconstruction of the {HOSVD} based canonical form of
  polytopic dynamic models.
\newblock In {\em 2007 International Symposium on Computational Intelligence
  and Intelligent Informatics}, pages 111--116. IEEE, 2007.

\bibitem{TD}
Ledyard~R Tucker.
\newblock Some mathematical notes on three-mode factor analysis.
\newblock {\em Psychometrika}, 31(3):279--311, 1966.

\bibitem{image}
M~Alex~O Vasilescu and Demetri Terzopoulos.
\newblock Multilinear analysis of image ensembles: Tensorfaces.
\newblock In {\em European Conference on Computer Vision}, pages 447--460.
  Springer, 2002.

\bibitem{tsvd}
Xiaoqiang Wang, Lejia Gu, Heung-wing~Joseph Lee, and Guofeng Zhang.
\newblock Quantum tensor singular value decomposition with applications to
  recommendation systems.
\newblock {\em arXiv preprint arXiv:1910.01262}, 2019.

\bibitem{QDT}
Nathan Wiebe, Daniel Braun, and Seth Lloyd.
\newblock Quantum algorithm for data fitting.
\newblock {\em Physical review letters}, 109(5):050505, 2012.

\bibitem{Z17}
Guofeng Zhang.
\newblock Dynamical analysis of quantum linear systems driven by multi-channel
  multi-photon states.
\newblock {\em Automatica}, 83:186--198, 2017.

\bibitem{ZNZ19}
Mengshi Zhang, Guyan Ni, and Guofeng Zhang.
\newblock Iterative methods for computing {U}-eigenvalues of non-symmetric
  complex tensors with application in quantum entanglement.
\newblock {\em Computational Optimization and Applications}, pages 1--20, 2019.

\bibitem{ZPR19}
Zhikuan Zhao, Alejandro Pozas-Kerstjens, Patrick Rebentrost, and Peter Wittek.
\newblock Bayesian deep learning on a quantum computer.
\newblock {\em Quantum Machine Intelligence}, 1(1-2):41--51, 2019.

\end{thebibliography}

\section{Author Contributions}
L.G., X.W., H.W.J.L and G.Z. developed the two quantum HOSVD algorithms and analyzed the computational complexity. L.G. and G.Z designed the hybrid quantum-classical recommendation method. L.G. summarized the results and wrote the manuscript. X.W., H.W.J.L and G.Z. reviewed and modified the manuscript.

\section{Competing Interests}
The authors declare that there are no competing interests.

\end{document}